\documentclass[hidelink]{aa}  

\usepackage{txfonts}
\usepackage{newtxtext,newtxmath}

\usepackage[T1]{fontenc}

\DeclareRobustCommand{\VAN}[3]{#2}
\let\VANthebibliography\thebibliography
\def\thebibliography{\DeclareRobustCommand{\VAN}[3]{##3}\VANthebibliography}

\usepackage{amssymb}	
\usepackage{float}
\usepackage{enumitem}

\usepackage{adjustbox}
\usepackage{graphicx,color,booktabs,tabularx}
\usepackage{multirow}
\usepackage{epsfig}
\usepackage{color}
\usepackage{setspace}
\usepackage{latexsym}
\usepackage{mathrsfs}
\usepackage{tikz}
\usepackage{rotating}
\usepackage{alltt}
\usepackage{longtable}
\usepackage{multicol}
\usepackage{tikz}
\usepackage{acronym}
\usepackage{pgfplots}
\usepackage{csquotes}
\usepackage{times}
\usepackage{placeins}
\usepackage{amsmath} 
\usepackage{hyperref}
\hypersetup{
    colorlinks=True,    
    urlcolor=blue,
    citecolor=blue,
}

\usepackage{geometry}

\begin{document} 

   \title{Ionised AGN outflows in the Goldfish galaxy $-$ The illuminating and interacting red quasar eFEDSJ091157.4+014327 at z$\;\sim\;0.6$ }

\titlerunning{Ionised outflow in a red quasar at z$\sim0.6$}

   \author{B. Musiimenta
          \inst{1,2}
          \and
          G. Speranza \inst{4,5}
          \and
          T. Urrutia\inst{3}
          \and
          M. Brusa\inst{1,2}
          \and 
          C. Ramos Almeida \inst{4,5}
            \and 
            M. Perna \inst{6}
            \and 
            I. E. López  \inst{1,2}
            \and 
            D. Alexander \inst{9}
            \and
            B. Laloux \inst{8,9}
            \and
            F. Shankar  \inst{11}
            \and
            A. Lapi  \inst{10}
            \and 
            M. Salvato \inst{7}
            \and 
            Y. Toba  \inst{12,13,14}
            \and 
            C. Andonie \inst{9}
            \and 
            I. M. Rodríguez \inst{8}
            }
          
   \institute{Dipartimento di Fisica e Astronomia 'Augusto Righi', Alma Mater Studiorum - Università di Bologna, Via Gobetti 93/2, 40129 Bologna, Italy
         \and
             INAF-Osservatorio di Astrofisica e Scienza dello Spazio via Gobetti 93/3, 40129 Bologna, Italy
             \and
             Leibniz-Institut für Astrophysik Potsdam, An der Sternwarte 16, 14482 Potsdam, Germany
             \and 
             Instituto de Astrofísica de Canarias, Calle Vía Láctea, s/n, 38205 La Laguna, Tenerife, Spain 
             \and 
             Departamento de Astrofísica, Universidad de La Laguna, 38206 La Laguna, Tenerife, Spain
             \and 
             Centro de Astrobiología (CAB), CSIC–INTA, Cra. de Ajalvir Km. 4, 28850 – Torrejón de Ardoz, Madrid, Spain.
             \and
             Max-Planck-Institut f\"ur extraterrestrische Physik, Giessenbachstra{\ss}e 1, D-85748 Garching bei M\"unchen, Germany
             \and Institute for Astronomy and Astrophysics, National Observatory of Athens, V. Paulou and I. Metaxa, 11532, Greece
             \and
             Centre for Extragalactic Astronomy, Department of Physics, Durham University, South Road, Durham, DH1 3LE, UK
             \and
             SISSA, Via Bonomea 265, 34136 Trieste, Italy
             \and
             School of Physics and Astronomy, University of Southampton, Highfield, SO17 1BJ, UK
             National Astronomical Observatory of Japan, 2-21-1 Osawa, Mitaka, Tokyo 181-8588, Japan
             \and
            Academia Sinica Institute of Astronomy and Astrophysics, 11F Astronomy-Mathematics Building, AS/NTU, No.1, Section 4, Roosevelt Road, Taipei 10617, Taiwan
            \and
            Research centrefor Space and Cosmic Evolution, Ehime University, 2-5 Bunkyo-cho, Matsuyama, Ehime 790-8577, Japan
     }

   \date{Received ...; accepted ...}

 
  \abstract
  {Evolutionary models suggest that the initial growth phases of active galactic nuclei (AGN) and their central supermassive black holes (SMBHs) are dust-enshrouded and characterised by jet or wind outflows that should gradually clear the interstellar medium (ISM) in the host by heating and/or expelling the surrounding gas. eFEDSJ091157.4+014327 (z$\sim$0.6) was selected from X-ray samples of eROSITA (extended ROentgen Survey with an Imaging Telescope Array) for its characteristics: red colours, X--ray obscuration (N$\rm{_H}\,=\,$2.7$\times$10$^{22}$ cm$^{-2}$) and luminous (L$_{\rm X}$=6.5$\times$10$^{44}$\,$\;\rm{erg\;s^{-1}}$), similar to those expected in quasars with outflows. It hosts an ionised outflow as revealed by a broad [O III]$\lambda 5007 \text{\AA}$ emission line in the SDSS integrated spectrum. For a proper characterisation of the outflow properties and their effects, we need spatially resolved information. }
   { 
   We aim to explore the environment around the red quasar, morphology of the [O III] gas and characterise the kinematics, mass outflow rates and energetics within the system. 
   }
  {We used spatially resolved spectroscopic data from Multi Unit Spectroscopic Explorer (MUSE) with an average seeing of 0.6" to construct flux, velocity and velocity dispersion maps. Thanks to the spatially resolved [O III]$\lambda 5007 \text{\AA}$ emission detected, we provide insights into the morphology and kinematics of the ionised gas and better estimates of the outflow properties.} 
  {We find that the quasar is embedded in an interacting and merging system with three other galaxies $\sim$ 50 kpc from its nucleus. Spatially resolved kinematics reveal that the quasar has extended ionised outflows of up to 9.2$^{+1.2}_{-0.4}$ kpc with positive and negative velocities up to 1000\,$\; \rm{km\;s^{-1}}$ and $-$1200\,$\; \rm{km\;s^{-1}}$, respectively. The velocity dispersion (W$_{80}$) ranges from 600 $-$ 1800\,$\; \rm{km\;s^{-1}}$. We associate the presence of high-velocity components with the outflow. The total mass outflow rate is estimated to be $\sim$10 M$_{\odot}$ yr$^{-1}$, a factor of $\sim$3 $-$ 7 higher than the previous findings for the same target and kinetic power of 2$\times$10$^{42}$\,$\;\rm{erg\;s^{-1}}$. Considering different AGN bolometric luminosities, the kinetic coupling efficiencies range from 0.01\% $-$ 0.03\% and the momentum boosts are $\sim$ 0.2.}
    {The kinetic coupling efficiency values are low, which indicates that the ionised outflow is not energetically relevant. These values don't align with the theoretical predictions of both radiation-pressure-driven outflows and energy-conserving mechanisms. However, note that our results are based only on the ionised phase while theoretical predictions are multi-phase. Moreover, the mass loading factor of $\sim$5 is an indication that these outflows are more likely AGN-driven than star formation-driven.  }


   \maketitle
%

\section{Introduction}
\label{introduction}

Since the discovery of supermassive black holes (SMBHs, 10$^6$- 10$^{10}$ $\rm{M_{\odot}}$) in the centres of galaxies \citep{lyndenbell1969,rees1984}, it has become clear that these SMBHs are closely connected to the formation and evolution of galaxies. There is a correlation between the properties of the host galaxy and the properties of the black hole for example the black hole mass and stellar velocity dispersion \citep{tremaine2002,kormendy2013,mcconnell2013,degraf2015}. It is believed that a mechanism connects the innermost regions, where the black hole's gravitational field is dominant to the larger scales. One proposed mechanism is the presence of outflows in the form of energetic jets or winds \citep{alexanderDM2012,harrison2017,harrisonCM2024}. These outflows are believed to play a crucial role in galaxy evolution by regulating the accretion onto and ejection of material from the SMBH \citep{kingp2015}. During this period called the 'feedback phase', the radiation pressure accelerates in the form of winds from the accretion disk in rapidly accreting sources into the interstellar medium (ISM), driving outflows and providing an efficient feedback mechanism \citep{silk1998,hopkins2009}. 

The outflows in active galactic nucleus (AGN) host galaxies have a complex nature, consisting of multiple phases ( \citealp[ionised, molecular and neutral;][]{Cicone2018,harrisonCM2024}). The ionised phase has been observed to extend over kilo-parsec scales in both local \citep{kakkad2022,speranza2022,toba2022,shen2023,venturi2023,zanchettin2023} and high redshift (z) AGN \citep{carniani2015,scholtz2020,cresci2023} using spatially resolved data. While the presence of ionised outflows can be detected by broad or shifted wings in the [O III]$\lambda 5007$$\text{\AA}$  emission line in integrated spectra \citep{mullaney2013,zakamska2016,toba2017}, high-resolution and high signal-to-noise ratio observations with integral field units (such as MUSE, KMOS and SINFONI) have played a crucial role in determining their extent (radius) and kinematics (velocity and geometry) \citep[e.g][]{scholtz2020,kakkad2022,kakkad2023,cresci2023}. These parameters are essential for calculating the mass outflow rates and energetics of the outflows. 

Theorists suggest that a short, luminous and dust-enshrouded phase plays a significant role in the accretion of SMBHs when their activity is at its peak (high Eddington ratio; \cite{hopkins2005,dimatteo2005,blecha2018}). The reddest quasars are known to have these characteristic high Eddington ratios \citep[][]{urrutia2012}. The dusty environments produce moderate obscuration and reddening that makes it difficult to identify these objects. The most effective way to identify these sources and gather information about nuclear obscuration and ongoing AGN activity is through mid-infrared (mid-IR) and large-area hard X-ray surveys, coupled with multi-wavelength data such as optical surveys. Previous studies have utilised this approach to select these rare sources to provide insights into AGN activity \citep[e.g.][]{perna2015,cresci2015,brusa2016,musiimenta2023} and others to explore their radio properties \citep[e.g.][]{klindt2019,fawcett2020}.

To obtain significant samples of obscured quasars, it is essential to have an all-sky X-ray survey. This need is being addressed by the eROSITA (extended ROentgen Survey with an Imaging Telescope Array) telescope \citep{predehl2021}, thanks to its large field of view, spectral resolution, and angular resolution. The first data release for the all-sky survey is now out (eRASS1, \citealt{merloni2024}).

To test the capabilities of the eROSITA survey, a mini-survey called eFEDS (eROSITA Final Equatorial-Depth Survey \citep{brunner2022,salvato2021}) was conducted during the performance and verification phase, covering $\sim$ 140 deg$^2$ 
of the sky in the Subaru Hyper-Supreme Cam SSP field and reaching the average depth of the planned all-sky survey. \cite{musiimenta2023} exploited the large eFEDS catalogue ($\sim11000$ AGN at z=0.5-3) by applying, for the first time simultaneously, a combination of 5 selection methods to isolate $\sim1400$ candidate AGN in the feedback phase. The five methods were based on optical and mid-IR colours and optical and X-ray spectral properties and they include;
\begin{itemize}
    \item[$\circ$] $\rm{{i - W3} > 4.6}$
    \item[$\circ$] $\rm{\left (r - W1> 4 \right) \land \left ( log \frac{F_{2-10 keV}}{F_{optical}} > 1 \right)} $
    \item[$\circ$]  $\rm{\left (r - W1> 4 \right) \land \left ( log \frac{F_{0.2-2.3 keV}}{F_{optical}} > 0.5 \right)} $
    \item[$\circ$] $\mathrm{\left (r - W1> 4 \right) \land (i - W4} > 7)$
    \item[$\circ$] ($\mathrm{N_{H}>10^{21.5} cm^{-2}}$) $\land$ ($\mathrm{\lambda_{Edd}}>$ effective Eddington limit for different values of $\mathrm{N_{H}}$).
\end{itemize}

Out of the selected candidates, only one source, eFEDS J091157.4+014327 (referred to as XID 439 in \cite{brusa2022} and ID 608 in \cite{salvato2021,musiimenta2023} and hereafter) has been selected based on all five different criteria. It is therefore, referred to as a red quasar based on the combination of optical and mid-IR colours \citep{musiimenta2023} see also \cite{urrutia2012,brusa2022}. The SDSS spectrum available for ID 608 (z=0.603) displayed a broad redshifted [O III]$\lambda 5007 \text{\AA}$ line ([O III] hereafter) with a Full Width at Half Maximum (FWHM) of approximately 1650\,$\; \rm{km\;s^{-1}}$. The observed luminosity of the [O III] emission (L$_{\rm{O III}}$) was measured to be around 2.6$\times$10$^{42}$ \,$\;\rm{erg\;s^{-1}}$, which is one to two orders of magnitude larger than what has been observed in local Seyfert galaxies. However, it is comparable to the [O III] luminosities of 1.26$\times 10^{43}$\,$\;\rm{erg\;s^{-1}}$ \citep{reyes2008} and  2.4$\times 10^{42}$\,$\;\rm{erg\;s^{-1}}$ (\href{http://research.iac.es/galeria/cristina.ramos.almeida/qsofeed/}{QSOFEED}) for a sample of optically selected QSO2s at 0.5$\rm{<z<}$0.83 and a subsample of type 2 quasars at z $\sim$0.14, respectively

ID 608 has also been previously reported in \cite{brusa2022} as an archetypal quasar in the feedback phase selected by eROSITA from the much smaller hard X-ray selected catalogue \citep{nandra2024}. From the X-ray point of view, ID 608 exhibits type 2 characteristics, with a luminosity of approximately 6.5$\times$10$^{44}$\,$\;\rm{erg\;s^{-1}}$ and a neutral hydrogen column density (N$\rm{_H}$) $\approx$ 2.75×10$^{22}$ cm$^{-2}$, when a simple power-law model is assumed. \citet{Waddell2023}, based on a comprehensive spectral analysis of the eFEDS hard X-ray sample, reported that ID 608 is best fit with an ionised warm absorber, supporting a scenario in which the outflowing ionised gas seen in the optical band may imprint its presence also in the X-ray spectrum. 
 The  Eddington ratio of ID 608 (L$_{\rm{bol}}$/L$\rm{_{Edd}}$) was derived in \cite{brusa2022} as $\sim$ 0.25 and revised later in \cite{musiimenta2023} as 0.99 using the black hole mass ($\log(\rm{M_{BH}})=8.4$) from \cite{shen2023}. These properties and the fact that it is the only candidate quasar in the feedback phase selected simultaneously from multiple diagnostics make ID 608 an ideal target for further studies on the spatially resolved analysis of ionised winds in obscured quasars at intermediate redshifts and their interaction with the surrounding environment.

This study aims to analyse the observational data for ID 608 from the Multi Unit Spectroscopic Explorer (MUSE) instrument. Our focus is to investigate the kinematics of the outflows observed in this particular system and the environment around it and shed light on the underlying mechanisms responsible for these outflows. This paper is organised as follows. 
Section \ref{MUSEobservations} describes the MUSE observations of our source and data reduction. Section \ref{methodology} describes methods and results concerning companion galaxies, kinematics, and ionised outflow properties. Section \ref{discussion} describes the interacting system discovered and mass outflow rates and energetics.  Conclusions are presented in Sect. \ref{conclusions}. Throughout the paper, we adopt the cosmological parameters $\rm{H_{0}=70 km\;s^{-1}Mpc^{-1}}$, $\rm{\Omega_{m}=0.3}$ and $\rm{\Omega_{\Lambda}=0.7 }$\citep{spergel2003}.

\section{MUSE observations and data reduction}
\label{MUSEobservations}

\begin{figure*}[!t]
\centering
    \includegraphics[width=\textwidth]{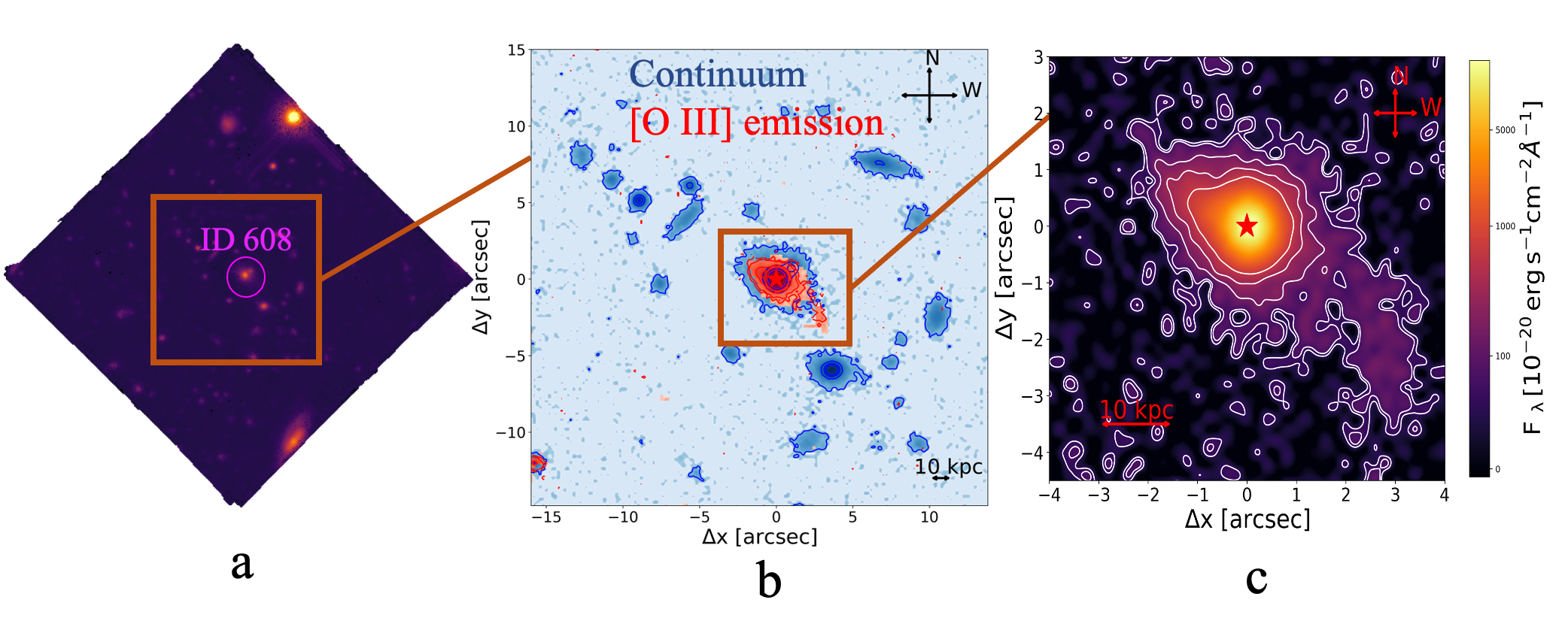} 
\caption{Full data cube over the whole wavelength range covered in our MUSE observation and the continuum and [O III] emission for size-reduced cubes. Panel a (left) is the image of the full MUSE's FOV ($1' \times 1' \cong$ 402 kpc $\times$ 402 kpc) obtained after data reduction. ID 608 is marked by a purple circle. Panel b (centre) shows an inner zoom of the cube ($30'' \times 30'' \cong$ 201 kpc $\times$ 201 kpc) with the continuum in blue, extracted from two regions of the spectrum (around [7825, 7900]$\text{\AA}$ and [8100, 8150]$\text{\AA}$ observed wavelengths) and the [O III] emission in red extracted from observed wavelengths 7890$\text{\AA}$ to 8100$\text{\AA}$. 
Panel c (right) shows a further zoom ($8'' \times 7.5'' \cong$ 53 kpc $\times$ 50 kpc) of the [O III] emission centred on the quasar position. The position of the nucleus of the quasar obtained from the maximum of the [OIII] emission is marked by the red star. The contours in both panels b and c represent 3$\sigma$, 5$\sigma$, 15$\sigma$, 30$\sigma$, and 60$\sigma$. North is up, East is left. }
\label{fullcube}
\end{figure*}

The MUSE instrument located at the Very Large Telescope (VLT) of the European Southern Observatory (ESO) combines the power of a wide-field imager and an integral field spectrograph allowing scholars to simultaneously obtain detailed images and spectra of galaxies. It operates in various optical wavelengths, capturing light across a wide spectrum from ultraviolet to near-infrared. Its high-resolution capabilities allow for detailed mapping of structures within galaxies revealing intricate details of phenomena such as AGN outflows and star-forming regions among others.

We conducted observations of ID 608 at RA(J2000); 09:11:57.557 and DEC(J2000); +01:43:27.54) using the MUSE instrument \citep{bacon2010} as part of  proposal ID:108.22M6.001. 
 Our observations spanned five nights; December 28$^{\rm{th}}$ 2021, December 29$^{\rm{th}}$ 2021, January 3$^{\rm{rd}}$ 2022, January 4$^{\rm{th}}$ 2022, and January 5$^{\rm{th}}$ 2022 in Wide Field Mode (WFM). These yielded an average seeing of 0.6" for a total observation time of 6.1 hours. 
The field of view is 1$\times$1 arcmin$^2$ with a pixel size of 0.2 arcsec. Our observations cover a spectral range of 4650–9300 $\text{\AA}$ sampled at intervals of 1.25 $\text{\AA}$ with a spectral resolution R $\approx$ 2800.


The ESO archive provides both raw and reduced data cubes that are ready for science 
 analysis. However, to ensure the highest data quality, we opted to work with the raw data and perform the data reduction ourselves following the manual from the AIP website\footnote{https://data.aip.de/data/musepipeline/v2.6/muse-pipeline-manual-2.6.pdf}. This process uses \texttt{EsoReflex} \citep{freudling2013} and allows us to have full control over the reduction process instead of relying on solely automatically reduced MUSE data cubes generated by the ESO pipeline.

For the raw data pre-processing, we utilised the MUSE data reduction software (Version 2.8 of the pipeline \cite{weilbacher2020}). We used all the files provided by the ESO archive for our observation, except for the line spread function (LSF) profile and geometry and astrometry tables. To ensure accuracy we calculated the LSF profile ourselves and substituted the geometry and astrometry tables with more recent ones obtained in December 2021, which were closer in time to our observation compared to the ones provided by ESO from April 2021.

The pre-processing steps included bias subtraction, dark calibration, flat-fielding, wavelength calibration, LSF calibration, and twilight calibration. Since our source was observed for five nights, we combined the preprocessed data into a single datacube using \texttt{muse\_exp\_align} and \texttt{muse\_exp\_combine} recipe as explained in the manual. The full cube collapsed over the entire wavelength range is shown in panel a of Fig.\,\ref{fullcube}, where the nucleus is marked with a red star. 

Since we are interested in ionised outflows traced by the [O III] emission that only extends over a small portion of the FOV, we created a smaller cube around the position of ID 608. Panel b and c of Fig.$\;$\ref{fullcube} show the reduced datacubes of sizes of $30'' \times 30''$ (201 $\times$ 201 kpc) and $8'' \times 7.5''$ (54 kpc $\times$ 50 kpc), respectively. 
This [O III] emission line falls within the wavelength range of 7890 $-$ 8100 $\text{\AA}$ at our source redshift z$=$0.6031. Upon examining the intensity map in panel c of Fig.\,\ref{fullcube}, we observe extended [O III] emission in the southwestern (SW) region of the source and, to a lesser extent, also in the northeastern (NE). The structure of this source's [O III] emission as seen in panel c resembles the shape of the Goldfish. Therefore, we name this quasar a `Goldfish' galaxy for future reference. 

\section{Methodology and results }
\label{methodology}

\subsection{Gas distribution and searching for companion galaxies}
\label{gasdistributionandcompanions}

\begin{figure*}[!h]
\centering
    \includegraphics[width=0.95\textwidth]{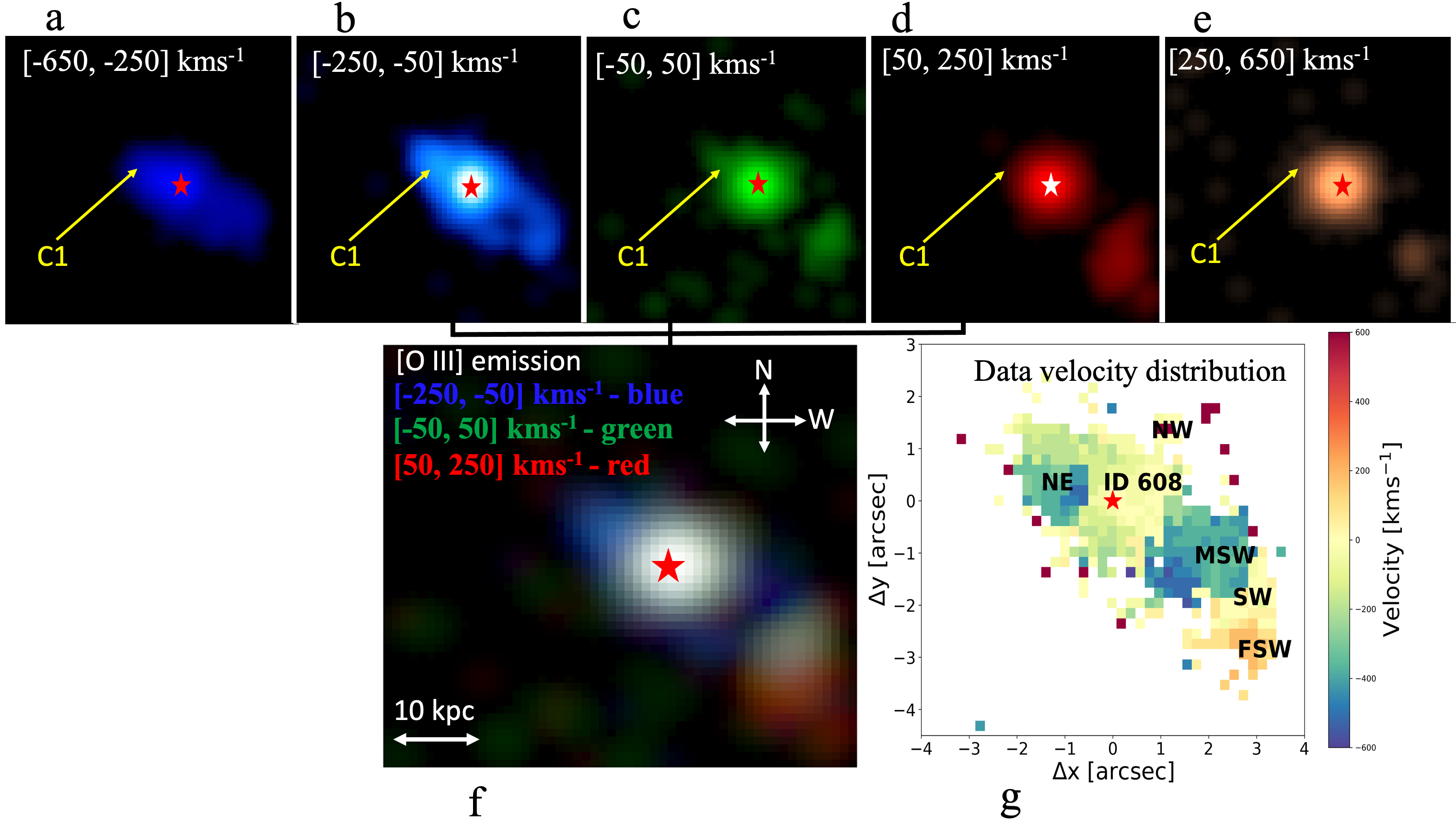} 
\caption{ RGB colour image, velocity channels and velocity distribution for the [O III] emission obtained from data without fitting. The different panels are named as a, b, c, d, e, f, and g. Panels a to e: [O III] intensity maps extracted in different velocity channels, as labelled. The arrows indicate the position of companion one (C1). Panel f: RGB colour image constructed from three [O III] velocity bins ([-250,-50] km s$^{-1}$, [-50,-50] km s$^{-1}$ and [50,250] km s$^{-1}$). Panel g shows the velocity distribution map attained without fitting to show the different regions as will be referred to in this study. The regions are NE, NW, SW, MSW, and FSW. The star shows the position of the nucleus of the quasar. The figures in all the panels are at the same scale and extracted from $8'' \times 7.5'' \cong$ 53 kpc $\times$ 50 kpc FOV. }
\label{velcolorimageconti}
\end{figure*}

\begin{figure*}[!h]
\centering
    \includegraphics[width=\textwidth]{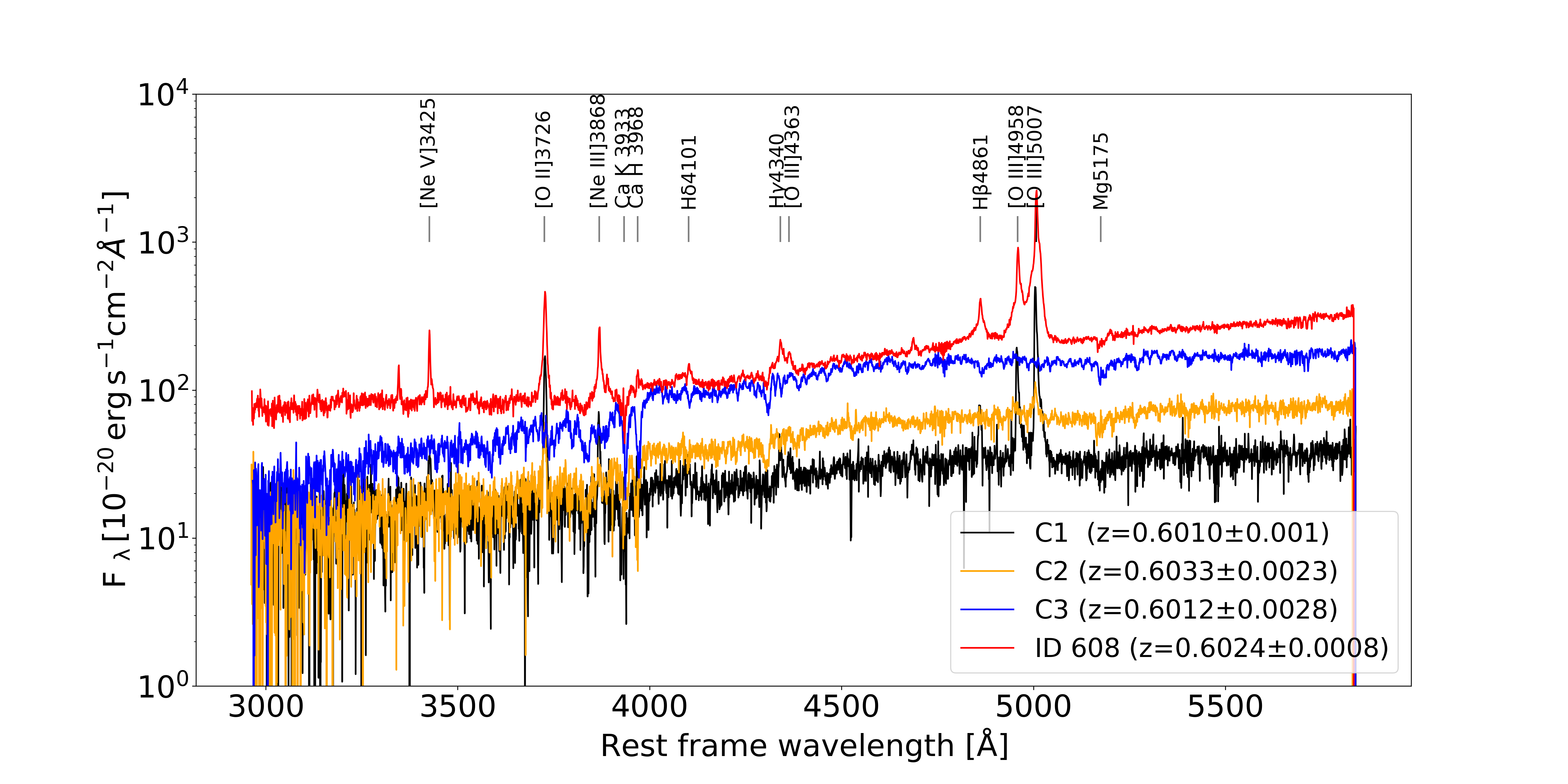} 
\caption{Spectra of the quasar and the companion galaxies C1, C2, and C3 with their estimated redshifts. The spectrum in red indicates the quasar (ID 608), companion galaxy one in black, companion galaxy two in yellow, and companion galaxy three in blue.}
\label{companions_spec}
\end{figure*}

To understand how the [O III] emitting gas is distributed within our system, we binned the data into different velocity channels.
The channels included velocities ranging from $-$650 to $-$250\,$\; \rm{km\;s^{-1}}$, $-$250 to $-$50\,$\; \rm{km\;s^{-1}}$ (the 'blue' channel), $-$50 to 50\,$\; \rm{km\;s^{-1}}$ (the 'systemic' channel), 50 to 250\,$\; \rm{km\;s^{-1}}$ (the 'red' channel), and 250 to 650\,$\; \rm{km\;s^{-1}}$.

We created flux maps or colour maps from these velocity channels, which are shown in panels a, b, c, d, e, and f of Fig.$\;$\ref{velcolorimageconti}. The top panels (a, b, c, d, and e) show all the velocity channels individually, while panel f shows a combined RGB image of the blue, systemic, and red channels. To determine the centroid of the quasar, we used a fixed wavelength (centroid) of 8026.5$\text{\AA}$ (corresponding to the [OIII]5007 emission line at z=0.6031). The centroid has been determined using \texttt{QFitsView} \citep{ott2012} by subtracting the continuum from the quasar nuclear spectrum extracted within an aperture of 0.6" $\times$ 0.6" (4$\times$4 kpc). The pixel with the maximum [O III] emission was considered the 'quasar nucleus' and is marked by a star in all the panels. Its wavelength with the maximum [O III] emission was used as the centroid of the quasar (quasar centroid hereafter).

 We generated a velocity map from the data without fitting, which involved extracting a spectrum for each pixel around the 7890-8100$\text{\AA}$ wavelength range and determining the wavelength corresponding to the peak of the [O III] emission line.
The velocity map (shown as 'data velocity distribution') in panel g of Fig.$\;$\ref{velcolorimageconti}, indicates the velocity distribution of the gas that is asymmetric with velocity gradients. It also shows the different regions as explained below.

In Panel a, the gas distribution shows an extended structure towards the south-west (SW) direction and a small extension towards the north-east (NE) direction. This SW extension is more visible in Panel b, where an additional extension in the NE appears as a second peak. This peak corresponds to higher velocities of up to $\sim$600\,$\; \rm{km\;s^{-1}}$ in the NE region of Panel f. Likely, this second peak located at 1.4$''$ or 9.4 kpc away from the nucleus of the quasar represents a companion galaxy or a merger (C1 in the figure, see further discussion below). The colour maps of velocity channels [-650,-250] km s$^{-1}$ and [$-$250, $-$50]\,$\; \rm{km\;s^{-1}}$ suggest the presence of a bubble-like structure in the middle SW position (MSW region in panel f and hereafter).

The systemic velocity, as shown in Panel c, is mainly centralised, with only a small portion showing up in the SW position (SW region in panel f and hereafter). Additionally, we observed the emission in the red velocity channel, shown in panel d, with most of it centralised near the quasar nucleus and extra emission in the SW region and far SW position (FSW region in panel g and hereafter). In Panel e, the velocity channel [250,650] km s$^{-1}$ mainly exhibits centralised emission near the quasar nucleus. Complete velocity channels from $-$1200 km\;s$^{-1}$ to 1200 km\;s$^{-1}$ at intervals of 100 km s$^{-1}$ are shown in Appendix \ref{Velocity_channels} with more visible extended structures.

To search for additional emitting sources, we analysed the spectra of the sources within the larger FOV closer to the quasar (201 $\times$ 201 kpc) and estimated their redshifts. Based on the presence of continuum emission and spectral lines (absorption or emission), in addition to C1 we discovered other two companion galaxies (C1 and C2) with redshifts consistent with that of the quasar. 

The spectra of ID608 and the three companion galaxies are shown in  Fig.~\ref{companions_spec}. 
The redshifts of these sources were estimated from different emission and absorption lines observed, that is [Ne V] 3425, [O II] 3726, Ca K and H, H$\delta$ 4101, H$\beta$ 4861, [O III] 5007, and Mg 5175. [Ne V] 3425, H$\beta$ and [O III] 5007 were only used if they appeared significant in the spectrum. All these lines are marked in Fig.~\ref{companions_spec}. C1 is found at a redshift of z $=$ 0.6010$\pm$0.001, while ID 608 is at z $=$ 0.6024$\pm$0.0008. Both C1 and ID 608 exhibit strong emission lines in their spectra. C2 on the other hand, displayed weak emission lines and is continuum-dominated with a redshift of z $=$ 0.6033$\pm$0.0023. In the case of C3, absorption lines were observed and it is also continuum-dominated with a redshift of z $=$ 0.6012$\pm$0.0028 displaying a significant 4000$\text{\AA}$ break. 

\begin{figure}[!h]
\centering
    \includegraphics[width=0.45\textwidth]{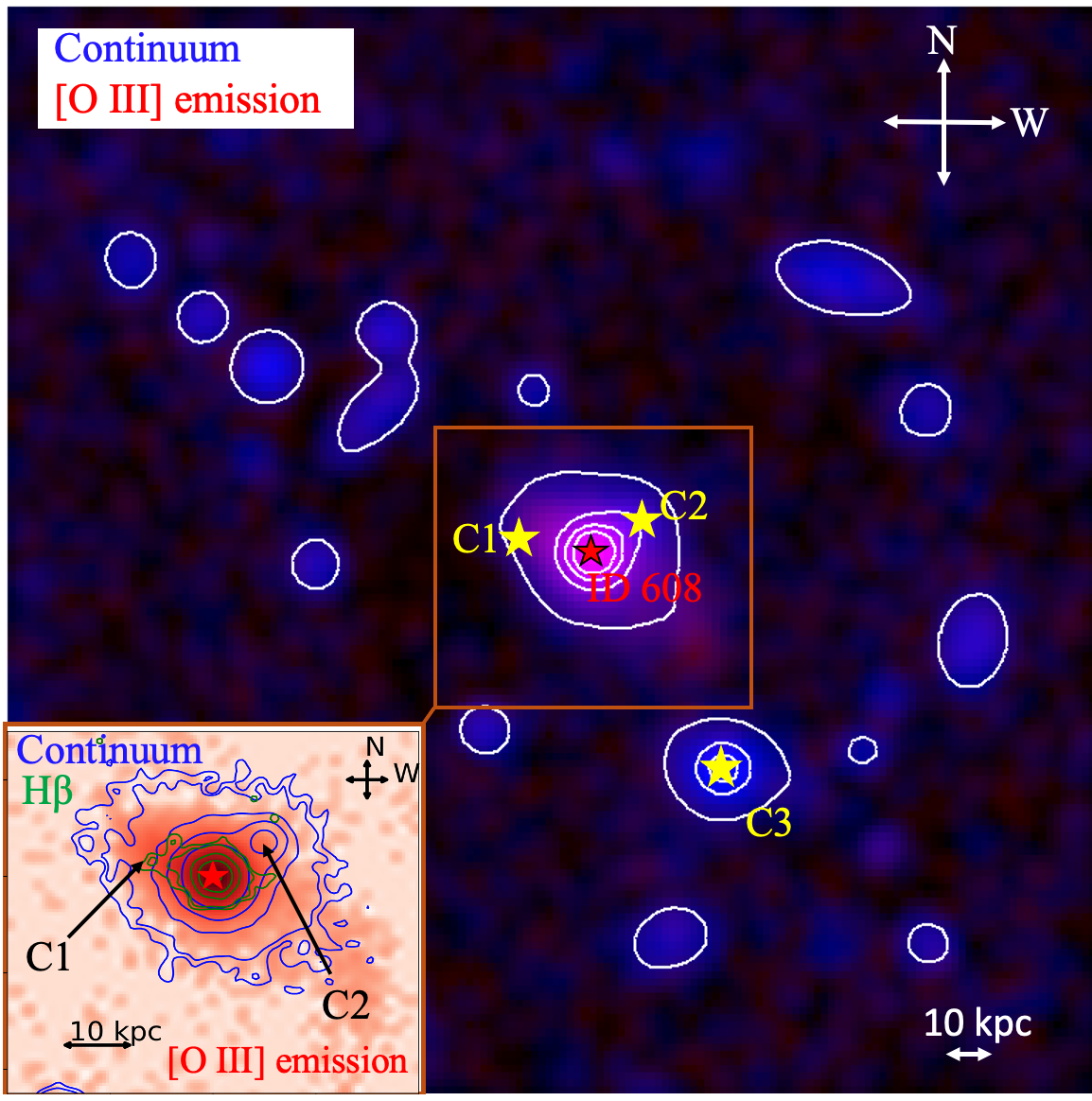} 
\caption{Colour map of the continuum emission overlapped with the [O III] emission in $30'' \times 30'' \cong$ 201 $\times$ 201 kpc FOV. The continuum emission is shown in blue with corresponding white contours and the [O III] in red (no contours). The companion galaxies are marked by yellow stars, while ID 608 nucleus is marked by a red star. In the bottom left of this figure, we show the zoom-in map of the continuum (blue contours), [O III] (red without contours), and H$\rm{\beta}$ (green contours) emission. The positions of C1 and C2 are indicated that do not appear to emit in H$\rm{\beta}$. The contours in the zoom in represent 3$\sigma$, 5$\sigma$, 15$\sigma$, 30$\sigma$, and 60$\sigma$.}
\label{companions}
\end{figure}

The three companions are marked as yellow stars in Fig.~\ref{companions}. C1 is at 1.4$''$ (9.4 projected kpc), C2 is at 1.3$''$ (8.7 projected kpc) and C3 is at 7$''$ (47 projected kpc) away from the quasar nucleus. The extended structure in the SW direction that is visible in the [O III] emission stretches towards the position of C3. In the bottom left corner of Fig.~\ref{companions}, we show the zoom-in map of the continuum, [O III], and H$\rm{\beta}$ emission around the quasar with positions of C1 and C2. From the zoom-in map, the peak of C2 is visually significant in the continuum emission. No continuum emission is seen in the MSW, SW, and FSW regions. C1 is slightly observed in H$\rm{\beta}$ while C2 is not, this is also seen in the spectra of the quasar and the companion galaxies in Fig.~\ref{companions_spec}. 

In summary, the analysis of the velocity distribution and the spectra suggests a complex system with various components. The presence of different companion galaxies at the same redshift as the quasar suggests the possibility of an ongoing merging of galaxies or galaxy interactions. Moreover, there is an extended structure (MSW, SW, and FSW) in the [O III] emission without continuum emission. This is likely to have resulted from an ongoing merger and/or interaction between C3 and the quasar.

\subsection{Analysis of the [O III] emission from different regions} 
\label{datafitting}

\begin{figure*}[!h]
\centering
    \includegraphics[width=\textwidth]{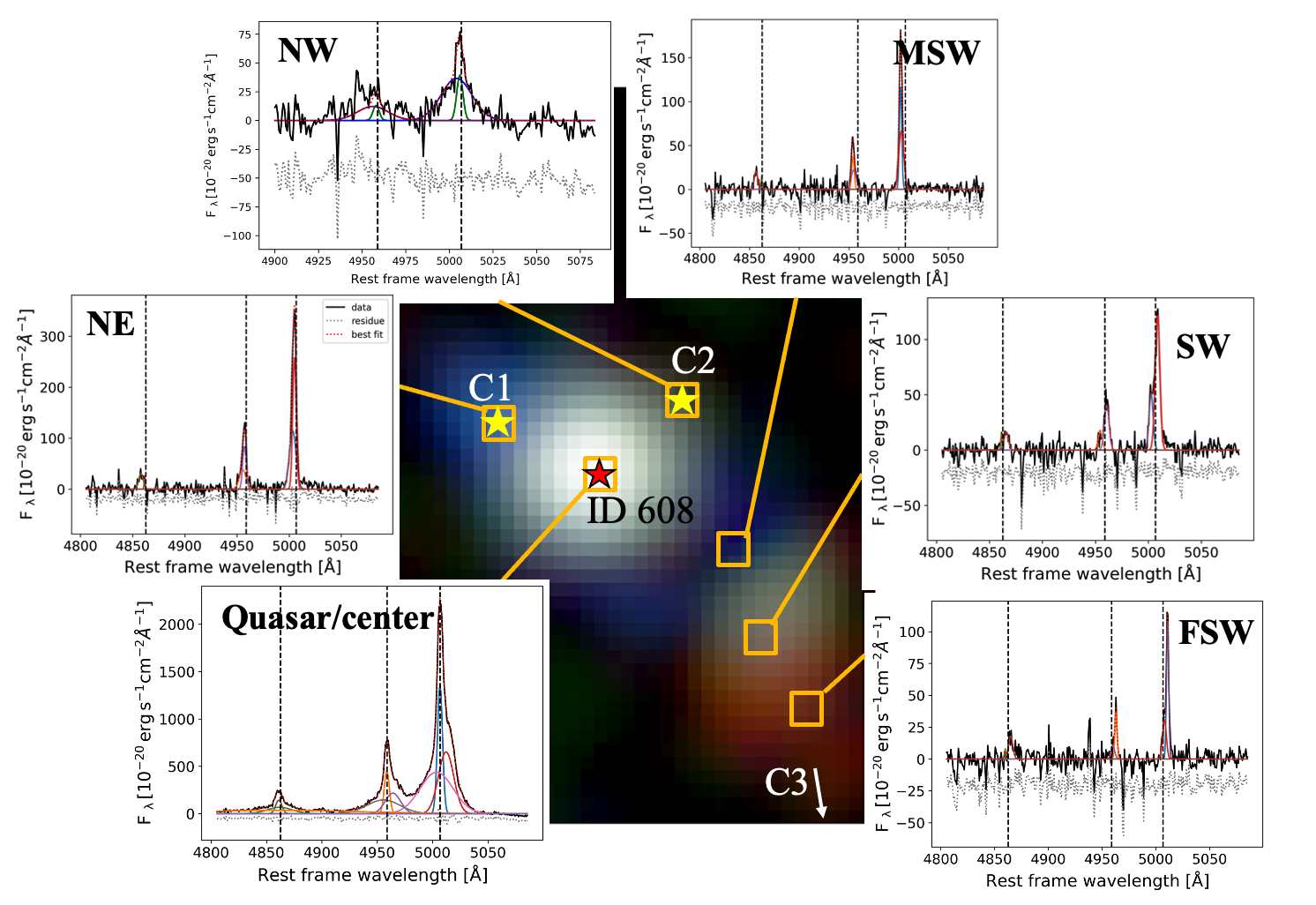} 
\caption{Spectra relative from the six regions shown in Fig.~\ref{velcolorimageconti}.
The spectra are extracted from a small area of a 0.6$'' \times 0.6''$ box at the given position in the FOV similar to the panel f of Fig. \ref{velcolorimageconti}. Within the six regions shown, only the nuclear region and NE are best fitted by more than two Gaussian functions. One or two Gaussian components can adequately describe all the remaining spectra. The legend for all the spectra is represented in the spectrum from the NE region where the black solid spectrum shows the data, the red dotted shows the best fit, the grey dotted spectrum shows the residual and all the other colours (blue, magenta, solid red, green) show the different components fitted. The three vertical lines highlight the centres of the H$\beta$$\lambda$4861, [O III]$\lambda$4949 and [O III]$\lambda$5007 emission lines. The yellow stars indicate the positions of C1 and C2. In the bottom right of the map, the white arrow points towards the position of C3 that is outside this map size. The different regions that is NE, NW, MSW, SW, FSW and ID 608 are indicated in the respective spectrum.}
\label{fluxandspectra}
\end{figure*}

We extracted spectra from a 0.6$'' \times 0.6''$ box from the regions ( NE, NW, SW, MSW, FSW and the centre) where we observed [O III] emission with velocity gradients. We fit the spectra with multi-Gaussian components (n=1-6) around  [O III]$\lambda \lambda 4949, 5007 \AA$ doublet using the \textit{lmfit} Python package and the in-house script described in \cite{speranza2023}. 
In this method, more than one Gaussian is fitted when the $\chi^{2}$ improves more than 10\% with respect to the previous fit. It is worth noting that the parameters of both emission lines were tied together and the flux ratios between the $\mathrm{[O III]\lambda{4959}}$ and $\mathrm{[O III]\lambda{5007}}$ lines were fixed at a ratio of 1:3 \citep{osterbrock1981}. The velocity dispersion of each kinematic component is left to vary. The different spectra are shown in Fig.$\;$\ref{fluxandspectra}.  

The spectrum from the NE region where an extra emission peak (C1) was observed shows an [O III] emission peak at a velocity offset of $\sim\;-235\;\rm{km \;s^{-1}}$ from the quasar centroid. We measured the line ratios to test the possibilities of AGN presence or quasar ionisation in the different regions and/or companion galaxies, if necessary. In the quasar, the line ratio log([O III]/H$_\beta$) $\sim$1. Since C1 is detected in both [O III] and H$_\beta$ emission, we measured the line ratio log([O III]/H$_\beta$) $\sim$0.9 which is compatible with the values that indicate the possible AGN ionisation. To check whether ID 608 is responsible for powering C1 or if there is a secondary AGN, we followed the method in \cite{Michele2023b}.  By assessing whether the line emissions in C1 can be attributed to the ionisation originating from ID 608, we calculated the ionisation luminosity using equation one (Eq. 1) in \cite{Michele2023b}. We obtained ionizing luminosity of $3\times 10^{43} \rm{erg\;s^{-1}}$, which is lower than the incident ionizing luminosity (see \cite{Michele2023b} for the formula) from ID 608 of $2\times 10^{45} \rm{erg\;s^{-1}}$. These values indicate that ID 608 serves as the ionizing source for C1 and exclude the possibility of the presence of a secondary AGN in C1.

The spectrum from the NW region (C2) shows a broad [O III] emission which could be due to the contamination from the quasar, though the continuum map shows significantly the presence of another source. Its [O III] emission peak observed is at a velocity offset of $\sim \; $-$93\;\rm{km \;s^{-1}}$ from the peak of the quasar. C2 is detected in [O III] but not in H$_\beta$. 
 
 We do not derive line ratios in C3 since it is not detected in [O III] and H$_\beta$. Its spectrum discussed in Sect.\;\ref{gasdistributionandcompanions} and Fig.\;\ref{companions_spec} is similar to that of an elliptical galaxy. 
 
The spectrum from the MSW region shows a blue shift in relation to the quasar centroid, while the spectrum from the SW region perfectly aligns with the quasar centroid. The spectrum from the FSW region exhibits a distinct red shift when compared to the quasar centroid. As seen in Fig.\;\ref{companions}, the H$_\beta$ detection in these regions is not significant because the Balmer line is faint (even in the centre). The [O III] extended emission in this extended structure is likely due to the interactions between the quasar and C3. The following section will explore the debate regarding whether these extended structures (MSW, SW and FSW) are indicative of outflows.  

\subsection{Spaxel-by-spaxel analysis of the [O III] emission}
\label{datafitting_spaxel}

\subsubsection{Is [O III] emission resolved?}
\label{psf}
\begin{figure*}[!h]
\centering
    \includegraphics[width=\textwidth]{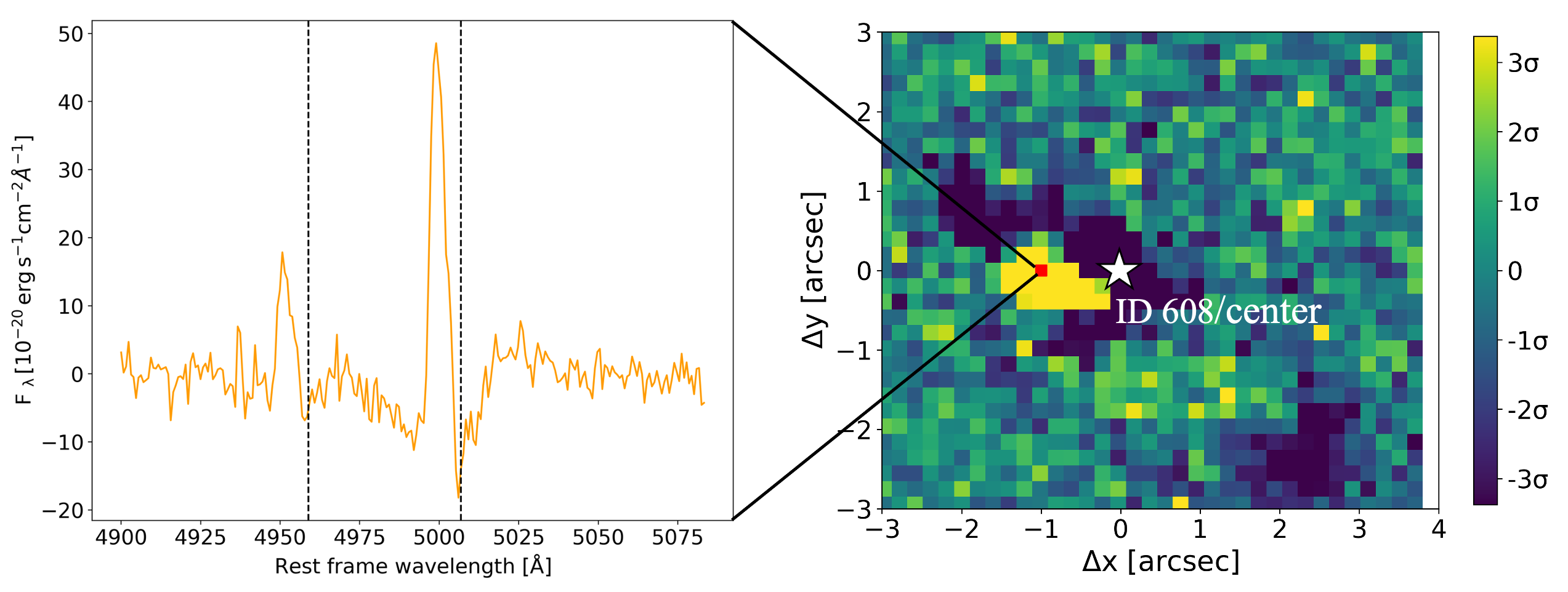} 
\caption{Residual spectrum and map after PSF subtraction show that the [O III] emission in our quasar is resolved. The left panel shows the residual spectrum analysis of a randomly selected pixel. The two vertical lines serve as markers highlighting the wavelengths of [O III]$\lambda$5007 and [O III]$\lambda$4949 emission lines for the centre of ID 608. The right panel showcases a residual map resulting from collapsing the residual spectrum of each pixel. The red square on the right panel corresponds to the specific pixel from which the residue spectrum was extracted in the left panel and the centre is marked by a white star.
}
\label{residue}
\end{figure*}

 Before performing a spaxel-by-spaxel [O III] fitting to produce kinematic maps, we followed the method in \cite{speranza2023} to determine whether the [O III] emission is spatially resolved or not, which is based on \cite{carniani2013} and also used in previous studies \cite[e.g.][]{carniani2015,kakkad2020a,kakkad2023}. This method known as 'point spread function (PSF) subtraction' allows us to assess the contribution of the PSF to the emission from the AGN. If the [O III] emission is resolved, the spectrum obtained from pixels away from the central region should differ from the spectrum obtained from the centre. For a complete description of this method, we refer the reader to Sect. 4.2 in \cite{speranza2023}. To perform PSF subtraction, we extracted a spectrum from five central pixels and fitted it with a multi-Gaussian model (see Sect. \ref{datafitting}) to obtain a 'PSF model'. Then, we extracted the [O III] spectrum for each pixel in the data cube and subtracted the PSF model. Before subtraction, a normalisation factor (considering the entire wavelength in the residual spectrum in Fig.$\;$\ref{residue} ) is applied to scale the amplitude of the PSF model. This scaling factor minimises the residues between the PSF model and the spectra from other pixels, as in \cite{speranza2023}(see also \cite{kakkad2023}). This process yielded a set of residual spectra for each pixel, as shown in the left panel of Fig.$\;$\ref{residue}. To visualise the residuals across the cube, we collapsed the residual spectra and created a residual map, displayed in the right panel of Fig.$\;$\ref{residue}. We observed excess emission at $>3\sigma$ near the central region in both the residual map and spectrum. This indicates that the [O III] emission is resolved. 

To properly measure the outflow energetics, it is important to account for the smearing effect caused by the spatial PSF, which can result in an overestimation of outflow measured values, particularly in objects where the nucleus outshines the extended emission \citep{husemann2016}. This effect can easily be addressed in unobscured quasars (e.g using QDeblend3D presented in \citealt{Husemann2013}, where the nuclear spectrum is scaled using the PSF modelled from broad line region (BLR) emission lines like H$\beta$ and subtracted from the spectrum of each off-nuclear spaxel).

We note that ID 608 is a type 2 (obscured) AGN. Modelling the emission from the (faint) BLR is therefore complex. In \cite{brusa2022} and \cite{musiimenta2023}, using the integrated SDSS spectrum, the best-fit model for the H$\beta$ complex included a narrow component, and an additional broad component with kinematics consistent with the ones observed in [O III] tracing the outflows, without any contribution from the unresolved BLR. The H$\beta$ extracted from the central 0.6$'' \times 0.6''$ box in the MUSE data cube is very faint (see the bottom left spectrum in Fig.~\ref{fluxandspectra}) and we do not detect any BLR component. 
 The PSF is therefore likely to go to zero in the spaxels very close to the centre, since it is constructed by scaling the nuclear spectrum to match the BLR wings in each spaxel. Therefore, no PSF deblending is required as BLR wings are undetected at radius, r $>$ 4 pixels.

In addition, we note that although the method removes emissions that can be ascribed to unresolved BLR, it may also wrongly remove flux from the components we are interested in, causing a severe underestimation of the outflow parameters. 
This is particularly important because the physical scale of our PSF is $\sim$4 kpc, large enough to contain most of the unsettled and outflowing gas. 

Given that we cannot detect clear emission from BLR and at the same time we do not want to disregard a significant portion of the outflowing flux just because it is unresolved, we will use the data cube without PSF subtraction for the subsequent analysis. At the end of Sect. \ref{outflowproperties}, we comment on the consequences of this assumption.

\subsubsection{Kinematics}
\label{kinematics_results}

After confirming that the [O III] emission is spatially resolved, we fit the [O III]$\lambda \lambda 4949, 5007 \text{\AA}$ doublet for each spaxel using multi-Gaussian components as described in Sect. \ref{datafitting}. We used the non-parametric approach where no physical meaning is ascribed to the single Gaussian components to avoid spaxel-by-spaxel parametric fitting that is too complex because of the many kinematic components that would be involved. These components are aimed at reproducing the line profile. By analysing the best-fit model obtained from this approach, we determined the kinematics of the gas around the quasar.

\begin{figure*}[h]
\centering
    \includegraphics[width=\textwidth]{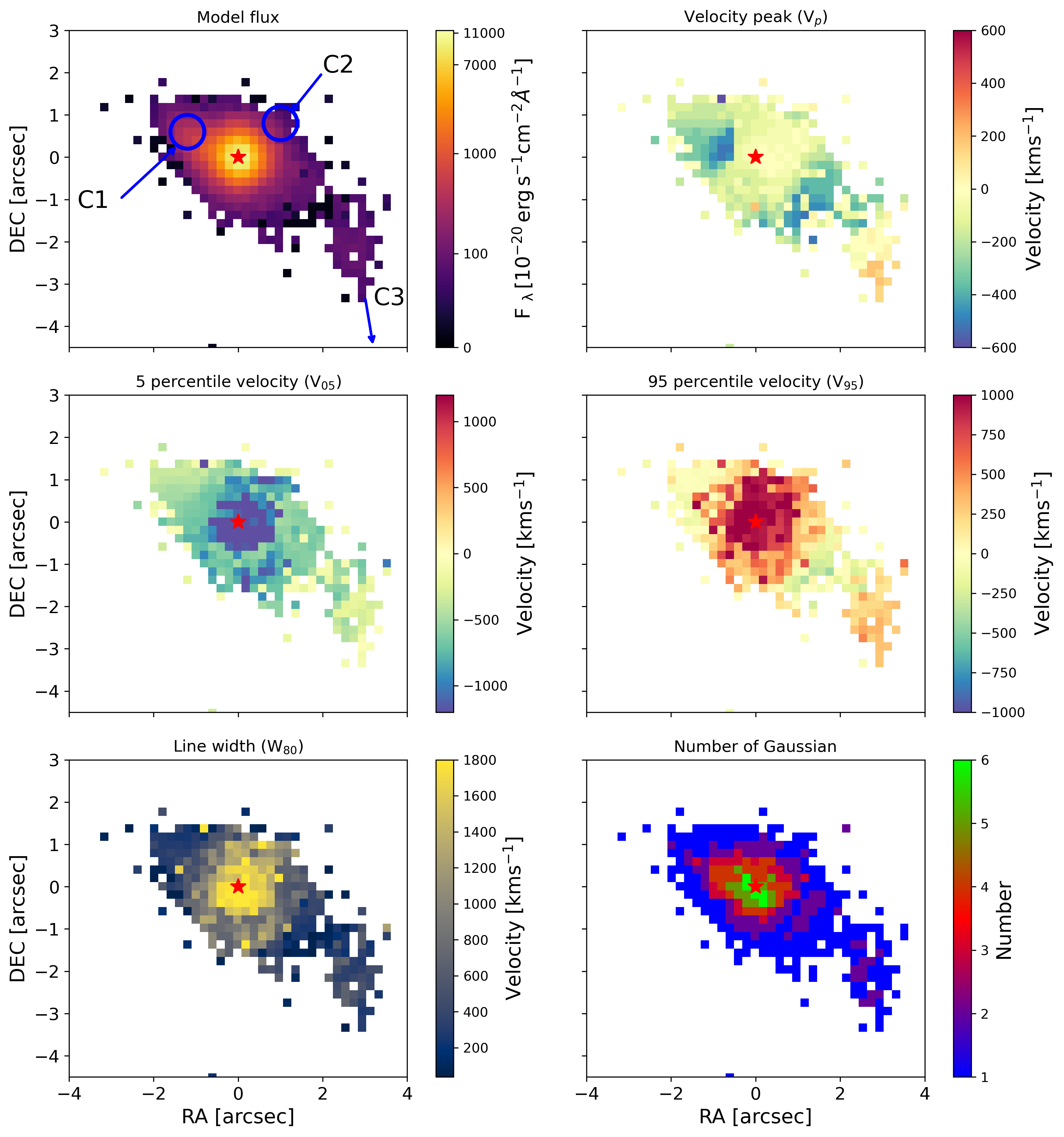} 
\caption{ Results from the spaxel-by-spaxel multi-Gaussian fitting of $8'' \times 7.5'' \cong$ 53 kpc $\times$ 50 kpc region around the quasar. Top left panel: the flux map. Top right panel: the velocity peak. In the middle panel, we present the V$_{\rm{05}}$ and V$_{\rm{95}}$ maps indicating the velocities at the 5$^{\rm{th}}$ and 95$^{\rm{th}}$ percentile, respectively. Bottom left panel: The line width (W$_{\rm{80}}$). Bottom right panel: the number of Gaussians fitted in each pixel. }
\label{velocitymaps_multi}
\end{figure*}

The kinematic maps are represented in Fig.$\;$\ref{velocitymaps_multi}. We define the five percentile velocity (V$_{\rm{05}}$) to trace outflow velocities of the approaching gas and 95 percentile velocity (V$_{\rm{95}}$) to trace outflow velocities of the receding gas. The line width that contains 80\% of the line emission flux that is, W$_{\rm{80}}$ = 90 percentile velocity $-$ 10 percentile velocity, is an approximation of the velocity dispersion. This velocity dispersion is equal to FWHM/2.35 for a Gaussian and tells us how turbulent the gas is. These parameters were obtained for each pixel resulting in the generation of velocity and flux maps in Fig.$\;$\ref{velocitymaps_multi}.

The flux map in the top left panel shows an extended emission in the NE and SW directions. The positions of C1 and C2 are indicated. C3 is outside this FOV, we indicate the direction of its position from the centre with the arrow. 

The velocity peak (V$_{\rm{p}}$) map in the top right panel displays systemic velocities close to the central region, followed by a sharp increase reaching $\sim$ $-$600\,$\; \rm{km\;s^{-1}}$ in the NE and MSW regions. In FSW, there is a rise in velocities to $\sim$ 200\,$\; \rm{km\;s^{-1}}$. This velocity peak map helps us visualise the kinematics of the ionised gas in our data: the pattern is asymmetric with a velocity gradient but no significant rotation is seen. The velocities in the extended structure in the SW direction are low and can be attributed to the ordinary movement of the gas than outflows. 

It is evident from the middle left panel that the quasar exhibits ionised outflows moving towards the observer, indicated by velocities (V$_{05}$) reaching up to a maximum of $-$1200\,$\; \rm{km\;s^{-1}}$. Within the spatial extension of up to $\sim$2" (13.4 kpc) away from the centre, these outflows have blue-shifted velocities greater than $-$500\,$\; \rm{km\;s^{-1}}$. On the other hand, the middle right panel shows receding gas (V$_{95}$) with velocities up to 1000\,$\; \rm{km\;s^{-1}}$, which is confined to a smaller region near the AGN, about 6.7 kpc in size for velocities greater than 500\,$\; \rm{km\;s^{-1}}$. These velocities greater than 500\,$\; \rm{km\;s^{-1}}$ are high enough to be associated with outflows. The red-shifted velocities observed in these kinematic maps were also identified from the SDSS integrated spectrum in \cite{brusa2022,musiimenta2023} with a velocity shift of 100\,$\; \rm{km\;s^{-1}}$ and 171\,$\; \rm{km\;s^{-1}}$.

In the bottom left panel, the line width is notably high ranging from 600 $-$ 1800\,$\; \rm{km\;s^{-1}}$ in regions where there is a high receding and approaching velocities. The high turbulence in these regions is associated with ionised outflows. As one moves farther away from the centre of the AGN, the velocity dispersion decreases. The velocity dispersion values in the MSW region are significantly lower than 500\,$\; \rm{km\;s^{-1}}$. Lastly, the bottom right panel displays the number of Gaussian fits performed in each pixel, providing further detail about the distribution and complexity of the observed data.

We tested a three-Gaussian fitting for n=1-3 with and without Voronoi binning. We found similar results (see Appendix \ref{kinematicsAP}).

\subsection{Ionised outflow properties}
\label{outflowproperties}

\begin{figure*}[!h]
\centering
\begin{tabular}{cc}
    \includegraphics[width=0.45\textwidth]{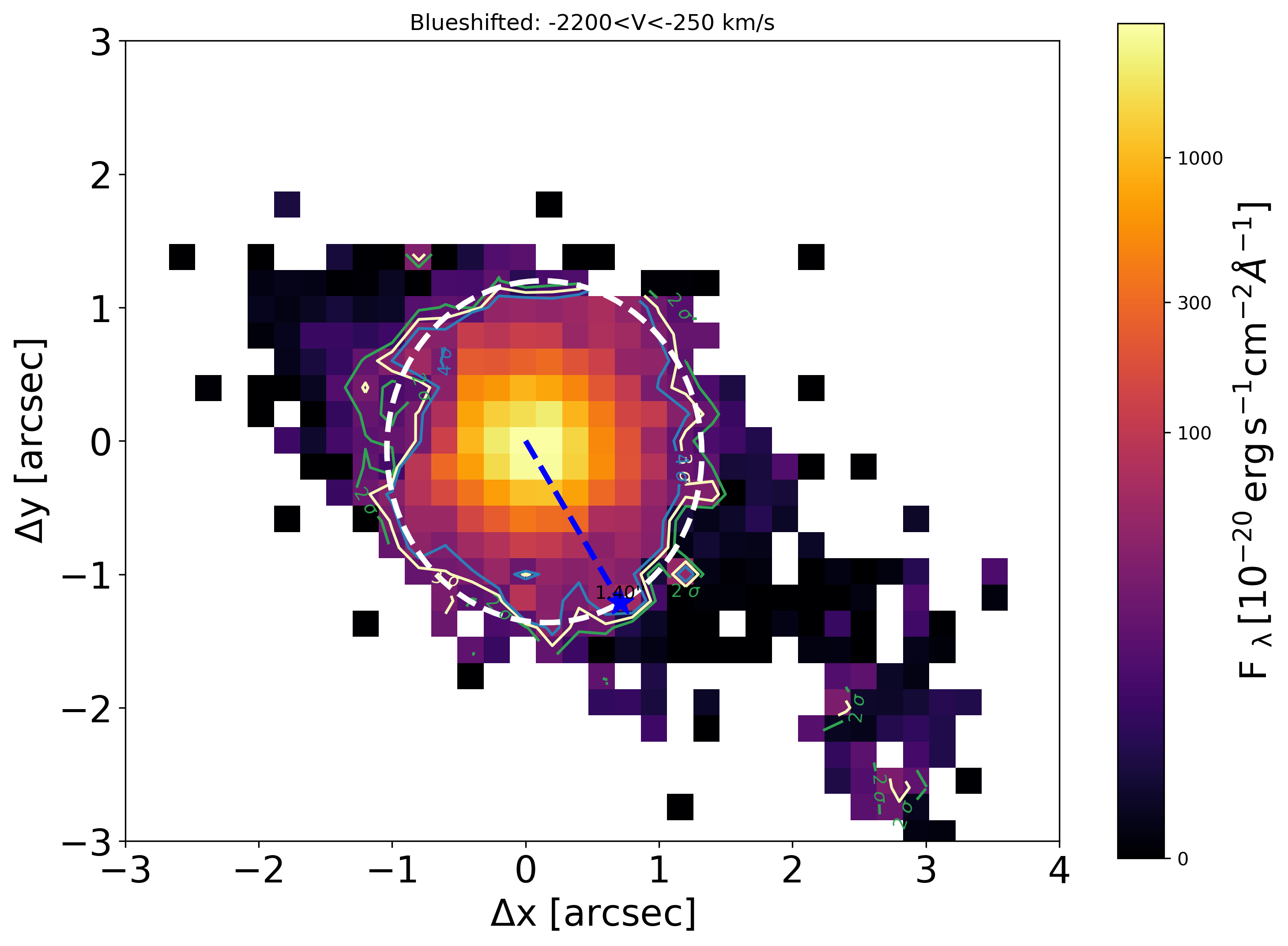} &
    \includegraphics[width=0.45\textwidth]{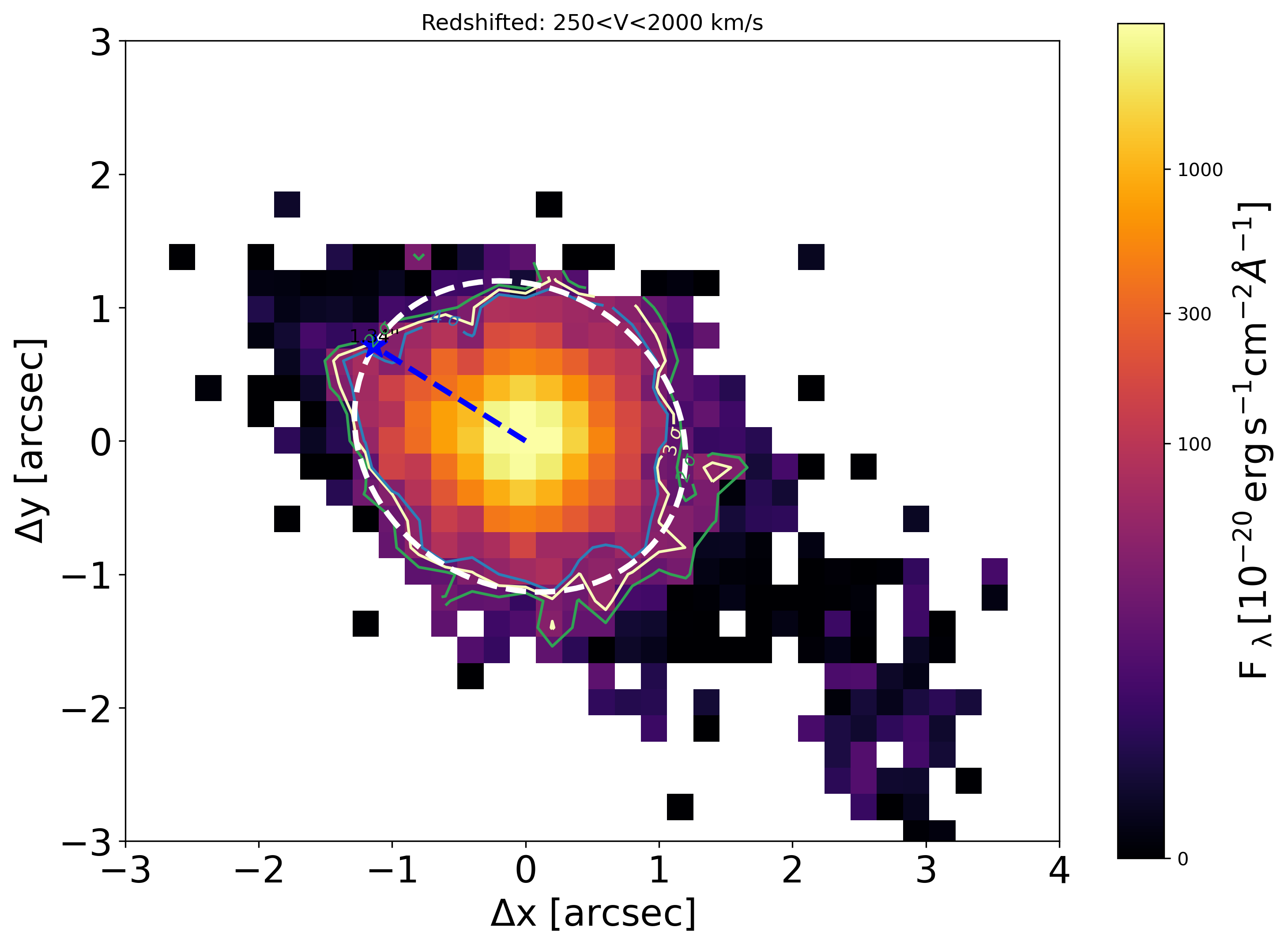} \\
\end{tabular}
\caption{ Flux maps for the blue-shifted and red-shifted velocities from the multi-Gaussian fitting. We select the region [-2200, $-$250]\,$\; \rm{km\;s^{-1}}$ to extract the flux map for negative velocities (the left panel). In the right panel, the flux map for the positive velocities  [250, 2000]\,$\; \rm{km\;s^{-1}}$. The light blue, yellow, and green contours indicate the 4, 3, and 2$\sigma$ levels, respectively. These contours help to determine the significance level of the measured flux values. The blue dashed line represents the maximum distance to the 3$\sigma$ fitted ellipse in white. }
\label{ellipsebluered}
\end{figure*}

In this section, we measure the ionised outflow properties including the mass of the outflow, the mass outflow rate and the kinetic power. To measure the outflow properties in this study, we construct flux maps for two different regions from the multi-Gaussian best-fit model described in Sect.\,\ref{datafitting_spaxel} that is, the region with blueshifted outflow velocities [$-$2200, $-250$] $\rm{km\;s^{-1}}$ (left panel of Fig.$\;$\ref{ellipsebluered}, blue region hereafter) and the region with redshifted outflow velocities [250, 2000] $\rm{km\;s^{-1}}$ (right panel of Fig.$\;$\ref{ellipsebluered}, red region hereafter). This allows us to isolate the flux that is only related to the outflow. The outflow properties are then measured for two separate regions and final values are later obtained by summing.

To measure the mass of the outflow, we employed the approach presented in \cite{carniani2015} using the formula,
\begin{equation}
    \label{mout}
   \mathrm{ M_{out}^{ion}}=\rm{5.33\times{10^7}M_{\odot}\left (\frac{C}{10^{[O/H]}}\right)\left(\frac{L^{ion}_{out}}{10^{44}erg s^{-1}}\right)\left(\frac{<n_{e}>}{10^3 cm^{-3}}\right)^{-1}},
\end{equation}
where $\rm{L^{ion}_{out}}$ is the outflow luminosity of the $\mathrm{[O III]}$ emission line in units of $\mathrm{10^{44}~erg\;s^{-1}}$, $\mathrm{n_{e}}$ is the outflowing gas electron density in units of $\mathrm{10^3  cm^{-3}}$,  $\mathrm{10^{[O/H]}}$ is the oxygen abundance in solar units and C is a factor that encodes the 'condensation factor' of the gas clouds (see \citealt[][for the details]{canodiaz2012}). The ionising gas clouds are always assumed to have the same density thus the condensation factor C is approximated to one and the metallicity of the outflowing material is always assumed solar. To determine the luminosity of the outflow ($\rm{L^{ion}_{out}}$), we follow the same approach outlined in \cite{speranza2023}. We fit an ellipse at 3$\sigma$ of the flux (white ellipse in Fig.$\;$\ref{ellipsebluered}) and measure the total flux within the ellipse as the flux associated with the outflow, for both the red and blue regions. The flux is corrected for dust extinction using the E(B-V) $=$ 0.63 value reported in \cite{kim2015} and relations presented in \cite{cardelli1989} and \cite{calzetti2000}. The extinction-corrected [O III] luminosity is 1.94$^{+0.02}_{-0.01}$$\times$10$^{42}$\,$\;\rm{erg\;s^{-1}}$  and 1.47$^{+0.78}_{-0.01}$$\times$10$^{42}$\,$\;\rm{erg\;s^{-1}}$ for red and blue regions, respectively, for a total of $\sim 3.4 \times 10^{42}$ erg s$^{-1}$. This value is comparable with the values reported in \cite{brusa2022} (2.6$\times$10$^{42}$\,$\;\rm{erg\;s^{-1}}$ ) and \cite{musiimenta2023} (2.2$\times$10$^{42}$\,$\;\rm{erg\;s^{-1}}$). The electron density can be measured directly using various methods such as the [SII]$\lambda\lambda$6716,30 line ratio, the auroral and transauroral line method and the ionisation parameter method (see \cite{davies2020} for detailed discussion of the methods). Unfortunately, the necessary emission lines for estimating the electron density are not available in our data. As a result, we assume an electron density (n$_{\rm{e}}$) of 500 cm$^{-3}$, consistent with the assumption made in \cite{brusa2022} also within the range of measured electron densities at z $\sim \;0.5\;-\;1.5$ \citep[e.g.][]{perna2015,perna2017b,cresci2023}. 
We calculate an ionised outflow mass of 15.5$\times$10$^{6}$ M$_{\odot}$ and 11.7$\times$10$^{6}$ M$_{\odot}$ for red and blue regions, respectively. The total outflow mass is 27$\times$10$^{6}$ M$_{\odot}$.

The spherical mass outflow rate can be deduced from Eq. \ref{mout} as
\begin{equation}
    \label{moutrate1}
    {\dot{\textrm{M}}_\mathrm{{out}}^{\rm{ion}}}= \rm{\Omega{\pi}R^2\rho{\nu} = \rm{3\frac{M_{out}^{ion}V_{out}}{R_{out}}}}, 
\end{equation}
where $\rm{V_{out}}$ is the outflow velocity, $\rm{R_{out}}$ (in kpc) is the radius at which the outflow is computed and $\mathrm{\rho=\frac{3M_{out}}{\Omega{\pi}R^3}}$ is the mean density of an outflow from the fluid field continuity equation with $\mathrm{\Omega{\pi}}$ as the solid angle covered by the outflow. Following \cite{speranza2023}, to measure the radius, we consider the maximum radius from the centre of the quasar to the 3$\sigma$ ellipse in Fig.$\;$\ref{ellipsebluered}. This procedure is carried out separately for the red and blue regions. To account for the seeing effect, we subtract half of the seeing FWHM (half width at half maximum) in quadrature from the measured radius following \cite{kangandwoo2018}. It is important to note that this correction only caters for the extension and does not account for the full smearing effect (see end of Section \ref{psf}). We find a radius of  $8.7^{+0.4}_{-0.8}$ kpc and $9.2^{+1.2}_{-0.4}$ kpc for red and blue regions, respectively. This value is comparable to 10 kpc in \cite{brusa2022} and \cite{musiimenta2023} where they assumed outflow extension as a half-light radius corresponding to the entire galaxy scale.

Regarding the measurement of the velocity, many methods have been used to measure the velocity for spatially resolved data \citep[see][for different velocity definitions and comparisons]{hervella2023}. We adopt as the velocity of the outflow (V$_{\rm{out}}$) the mean value between V$_{\rm{05}}$ and V$_{\rm{95}}$. We find the mean value, V$_{\rm{out}}$ $=$ 807.2\,$\; \rm{km\;s^{-1}}$ (red region) and 772.6\,$\; \rm{km\;s^{-1}}$ (blue region). The values of V$_{\rm{05}}$ and V$_{\rm{95}}$ used to calculate V$_{\rm{out}}$ are shown in Table \ref{tab:results_vmean}. We measure mass outflow rates of $\mathrm{\sim 5.9\;M_{\odot}~yr^{-1}}$ (red region) and $\mathrm{\sim 4.0\;M_{\odot}~yr^{-1}}$ (blue region). The total ionised outflow mass rate is 10$\;\rm{ M_{\odot}~yr^{-1}}$.

The corresponding kinetic power of the ionised outflow is estimated using
\begin{equation}
    \label{ke}
    {\dot{\textrm{E}}_\mathrm{{kin}}}={\frac{1}{2}\times{ {\dot{\textrm{M}}_\mathrm{{out}}^{\rm{ion}}}\times{\rm{V^{2}_{out}}}}}.
\end{equation}
This gives $\dot{\textrm{E}}_\mathrm{{kin}}\;\sim \;1\times$10$^{42}$\,$\;\rm{erg\;s^{-1}}$ (red region), $\dot{\textrm{E}}_\mathrm{{kin}}\;\sim\;1\times$10$^{42}$\,$\;\rm{erg\;s^{-1}}$ (blue region) and total $\dot{\textrm{E}}_\mathrm{{kin}}\;=\;2\times$10$^{42}$\,$\;\rm{erg\;s^{-1}}$.

The kinetic coupling efficiencies are defined as
\begin{equation}
    \label{kce}
    \xi={\frac{\dot{\rm{E}}_\mathrm{{kin}}}{\rm{L_{bol}}}},
\end{equation}
where $\rm{L_{bol}}$ is the AGN bolometric luminosity. $\xi$ was estimated based on the AGN bolometric luminosities obtained from \cite{brusa2022} (7.8$\times$10$^{45}$\,$\;\rm{erg\;s^{-1}}$, LB22 hereafter) and \cite{musiimenta2023} (1.6$\times$10$^{46}$\,$\;\rm{erg\;s^{-1}}$, LM23 hereafter). We obtain total kinetic coupling efficiencies of $\xi_{\rm{LM23}}\;=\;0.01$\% and $\xi_{\rm{LB22}}\;=\;0.03$\%. 

The ionised outflow momentum rate is given by
\begin{equation}
    \label{mr}
    \dot{\rm{P}}_{\rm{out}}={\dot{\rm{M}}_\mathrm{{out}}^{\rm{ion}}} \times \rm{V_{\rm{out}}}.
\end{equation}
The total value of $\dot{\rm{P}}_{\rm{out}}$ is determined to be 4.9$\times$10$^{34}$ $\rm{g\;cm\;s^{-2}}$. The rate of AGN momentum output is obtained through $\dot{\rm{P}}_{\rm{AGN}}$\;=\;L$_{\rm{bol}/c}$, resulting in values of 2.6$\times$10$^{35}$ $\rm{g\;cm\;s^{-2}}$ (LB22)  and 5.3$\times$10$^{35}$ $\rm{g\;cm\;s^{-2}}$ (LM23). The ratio $\dot{\rm{P}}_{\rm{out}}$/$\dot{\rm{P}}_{\rm{AGN}}$ is $\sim$0.2 for both luminosities (LB22 and LM23) and outflow regions (red and blue). The results showing the different values for both red and blue regions and the total (where applicable) are summarised in Table \;\ref{tab:results_vmean}. 

\begin{table}[htbp]
    \centering
    \caption{Ionised outflow properties for red region ([250,2000] km $\rm{s^{-1}}$) and blue region ([-2200,-250] km $\rm{s^{-1}}$) obtained using V$_{05}$ and V$_{95}$. Some of the values are reported in the main text considering the fact that the outflow properties have order of magnitude errors due to the assumptions involved in computing them.  }
    \label{tab:results_vmean}
    \begin{tabular}{l l l l}
        \toprule
        
         & \textbf{[250,2000]} & \textbf{[-2200,-250] } & \textbf{total}\\ 
        \midrule
         R$_{\rm{out}}$ [kpc] &  8.7$^{+0.4}_{-0.8}$ &  9.2$^{+1.2}_{-0.4}$ & $-$\\ 
         \addlinespace
         F$_{\rm{out}}$ [$10^{-14}$erg s$^{-1}$ cm $^{-2}$] & 1.29$^{+0.01}_{-0.01}$ & 0.97$^{+0.04}_{-0.04}$  & $-$\\
         \addlinespace
         $\rm{L^{ion}_{out}}$ [$ 10^{42}$erg s$^{-1}$]&  1.94$^{+0.02}_{-0.01}$ &  1.47$^{+0.78}_{-0.01}$ & $-$\\
         \addlinespace
         V$_{\rm{05}}$ [ km $\rm{s^{-1}}$] &  $-$952.9$^{+60.9}_{-34.4}$ & $-$915.7$^{+23.6}_{-20.8}$ & $-$\\
         \addlinespace
         V$_{\rm{95}}$ [ km $\rm{s^{-1}}$]&  661.5$^{+38.7}_{-40.2}$ & 629.4$^{+6.6}_{-21.5}$ & $-$\\
         \addlinespace
         M$^{\rm{ion}}_{\rm{out}}$ [$ 10^{6}$ M$_{\odot}$] &  15.5&  11.7 &  27.2 \\
         $\dot{\rm{M}}^{\rm{ion}}_{\rm{out}}$ [M$_{\odot}$ yr$^{-1}$]&   5.9 & 4.0 &  9.9 \\
         $\dot{\rm{E}}_{\rm{kin}}$ [$10^{42}$erg s$^{-1}$] &   1.2 &  0.8 & 2.0\\
         MLF &   2.9 & 2.0  &  4.9\\
         $\xi _{LM23}$ [ \%] &  0.008 & 0.005 &  0.013 \\
         $\xi _{LB22}$ [\%] &  0.016 & 0.010  & 0.026  \\
        $\dot{\rm{P}}_{\rm{out}}$ [$ 10^{34}$ g cm $s^{-2}$]&  3.0 & 2.0  & 5.0  \\
        $\dot{\rm{P}}_{\rm{AGN,LB22}}$ [$ 10^{35}$ g cm $s^{-2}$] & $-$  & $-$  & 2.6  \\
        $\dot{\rm{P}}_{\rm{AGN,LM23}}$ [$ 10^{35}$ g cm $s^{-2}$]& $-$  & $-$ & 5.3   \\
        \bottomrule
    \end{tabular}
\end{table}

There are several assumptions on different parameters such as geometry, n$_{e}$, temperature, the velocity of the outflowing gas and radius involved in computing outflow properties. These have been explored in several studies \citep[e.g,][]{husemann2016,fiore2017,davies2020,musiimenta2023,hervella2023,speranza2023}. The largest uncertainty is due to the assumptions made on n$_{e}$ that can change results by a couple of orders of magnitude as discussed in detail in \cite{davies2020}. Given that we assumed n$_{e}$ = 500 cm$^{-3}$, our results on the outflow properties can be considered as upper limits. Similarly, as discussed in \cite{hervella2023}, using different methods to measure the outflow velocity gives different results. For example, they reported higher values of mass outflow rates from parametric analysis using maximum outflow velocity than those derived from non-parametric analysis. Overall the values of outflow energetics in this study may have up to two orders of magnitude uncertainties due to the assumptions applied.

Finally, a comparison of outflow properties derived with and without PSF deblending for the few sources in \cite{husemann2016} that have similar characteristics as our quasar (very small or low H$\beta$, spatial resolution, L [OIII] luminosity and redshift) showed that the mass outflow rate obtained from the PSF subtracted datacube was generally only a factor of 2-5 lower than those obtained from the total flux datacube.

\section{Discussions}
\label{discussion}

\subsection{Interactions or merging in ID 608}
\label{morphology}

The colour maps discussed in Sect. \ref{gasdistributionandcompanions} and the kinematic analysis indicate that we may be dealing with a complex system with multiple dynamics occurring simultaneously. In Sect. \ref{gasdistributionandcompanions}, we present three galaxies (C1, C2, and C3) clearly associated with our quasar within $\sim$50 kpc from its nucleus. While C1 and C2 are both within 10 projected kpc from the nucleus (9.4 and 8.7 projected kpc away, respectively), C3 is instead at a farther distance of 47 projected kpc. 
The structure seen in the MSW, SW and FSW region extends from the quasar towards the position of C3 and appears to connect these two galaxies. 

\begin{figure}[!h]
\centering
    \includegraphics[width=0.45\textwidth]{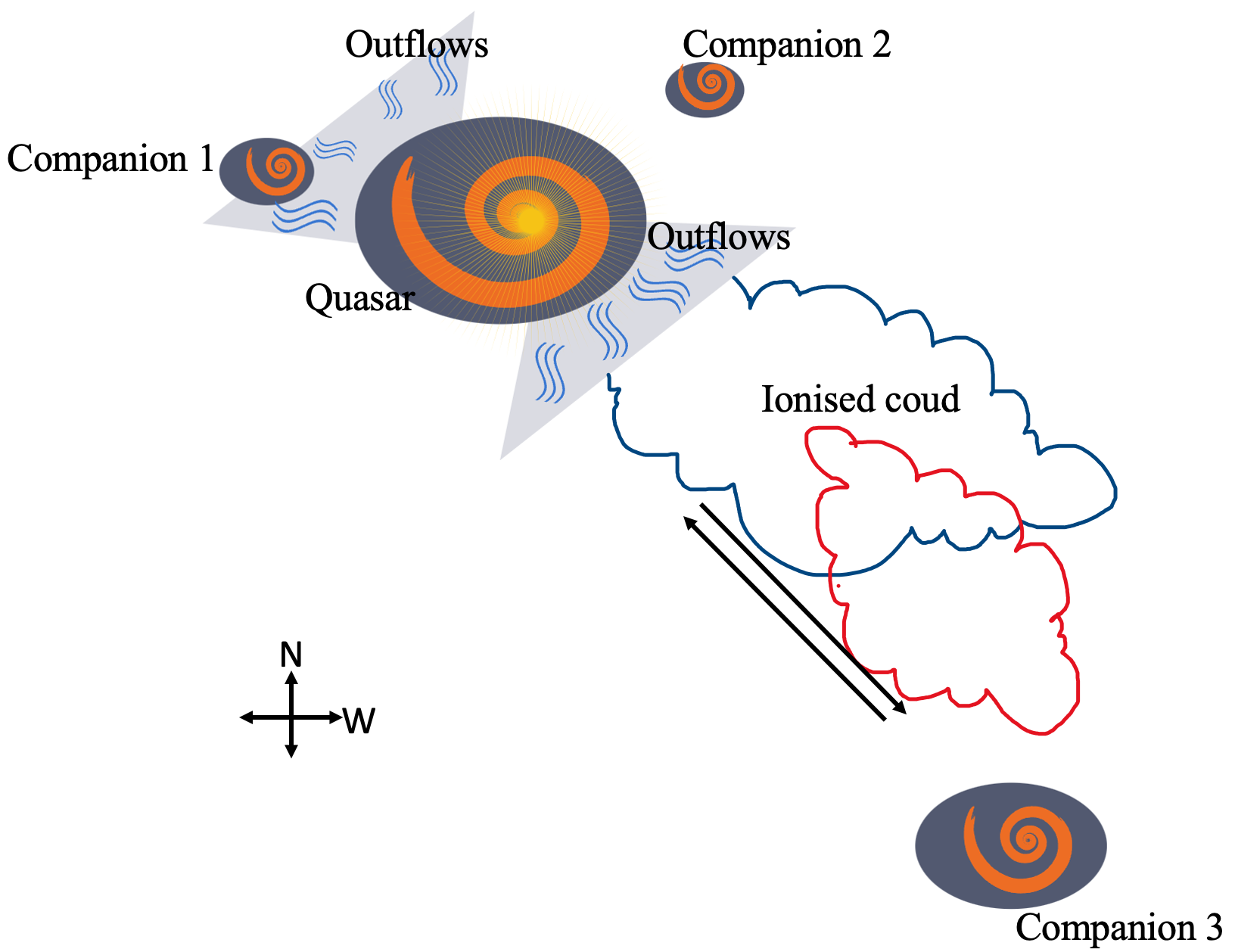} 
\caption{
Cartoon showing the quasar (ID 608) and the deductions we have made based on analysis of the data, velocity maps and flux maps. The companion galaxies are marked. The outflows are shown in the northeast (NE) and southwest (SW) directions, which we have denoted on the diagram. The extended structure is indicated as an ionised cloud in red and blue colours that indicate the receding and approaching movement of the cloud as seen in Fig. \ref{velcolorimageconti}. The possible interaction between the quasar and C3 is denoted by the black arrows. The distances are not to scale. There is no meaning attached to the sizes of the galaxies and outflows. 
}
\label{cartoon}
\end{figure}

Although we are limited by the available data to establish the physical connection between these companion galaxies and ID 608, we hypothesise that ID 608 is embedded in an interacting system with three other galaxies due to the following reasons: 1) The extended structure in the MSW, SW and FSW region has no corresponding continuum emission and it is likely a cloud of ionised gas traced by the  [O III] emission; 2) based on the kinematic analysis (see Sect. \ref{kinematics_results}), this extended structure has low-velocity dispersion (W$_{80}\;<\;600$ km $\rm{s^{-1}}$) and corresponding low velocities (< 500 km $\rm{s^{-1}}$) which does not align with the presence of outflows. Therefore, this extended structure could be an 'ionised cloud' of gas in the form of a tidal tail which formed as a result of merging or gravitational interactions (post and ongoing) between ID 608 and C3. Moreover, there is gas rotation going on in this structure as seen in the colour images and the velocity peak map (with velocity gradient) which confirms our hypothesis. It seems that C3 may be fated to merge with the quasar or we are seeing a post interaction.  This is a common feature of luminous red quasars showing prominent outflow signatures (e.g. \citealt{Wylezalek_2022}). The right panel of Fig.$\;$\ref{cartoon} visually depicts this ongoing interaction.

This discovery points to the fact that mergers trigger the growth of SMBH \citep{hopkins2008}. Moreover, the morphology of this system explains the observed properties of quasars with outflows which include optical and X-ray obscuration and red colours. The complexity discovered in this obscured red quasar also highlights the importance of X-ray selection methods to select these kinds of objects that are buried in interacting systems.

\subsection{Comparison of outflow properties with theoretical predictions}
\label{mdot}
The mass outflow rates of ID 608 have been previously studied using the SDSS integrated spectrum \citep{brusa2022,musiimenta2023}. Using spatially resolved analysis, we estimated the mass outflow rates and energetics of ID 608 as presented in Sect. \ref{outflowproperties}. We obtained total outflow mass of 27$\times$10$^{6}$ M$_{\odot}$, ionised outflow mass rate of 10  M$_{\odot}$ yr$^{-1}$,  kinetic coupling efficiencies of $\sim$ 0.01\% to 0.03\%. The momentum boosts are $\sim$0.2.

The mass outflow rate values obtained in this study using our spatially resolved data are generally higher than the values reported in \cite{musiimenta2023} ($\mathrm{\sim 2.6~  M_{\odot}~yr^{-1}}$), derived under similar assumptions on n$_e$, T and geometry. This is because the methodology used to compute outflow properties such as luminosity, radius and velocity differ from the ones assumed previously and can bring discrepancies in the final values obtained. Moreover, the spatially resolved information available in this study, allowed us to compute a more reliable mass outflow rate by summing the different regions of the outflow instead of assuming only the redshifted broad component of the integrated spectrum. This highlights the power of spatially resolved spectroscopy to better characterise the kinematics and extension of the outflowing gas.

Our results are comparable to other ionised outflow properties in previous studies of obscured AGN at $\rm{z\;\sim\;0.1}$ that have been computed using spatially resolved data. For example in \cite{speranza2023} where a similar method (see Sect. \ref{outflowproperties} ) was used to compute the outflow properties, for a sample of four local type 2 quasars with spatially resolved outflows, they obtained ionised mass outflow rates of 3.3 $-$ 6.5 M$_{\odot}$ yr$^{-1}$ within 3.1 to 12.6 kpc which are less than the values obtained in this study but within the same extension radius. 
Our values are within the ranges of 0.1 $-$ 23 M$_{\odot}$ yr$^{-1}$ as ionised outflows in \cite{musiimenta2023} at redshift z $\sim$ 0.5 $-$ 1 and with the same AGN bolometric luminosity range as our quasar.

\cite{musiimenta2023} reported a strong correlation between the mass outflow rate and the AGN bolometric luminosity, holding also for sources that show high outflow velocity but relatively low bolometric luminosities. The revised values of outflow parameters for ID 608 presented in this work are still fully consistent with the proposed correlation. ID 608 has high outflow velocity with a corresponding low ionised gas mass, 
further supporting the hypothesis of the presence of a fast-moving, less massive wind characteristic of the initial stages of the blow-out phase. 

Moreover, as described in Sect. \ref{morphology}, ID 608 is possibly going through early-stage merger activity and was selected based on its red colours, X-ray brightness and obscuration. This indicates that these selection methods can be used to effectively isolate objects in the initial stages of outflowing phases that is, objects buried in dust enshrouded environments.

Although the direct comparison of observations results to theoretical prediction is not yet reliable, we note that our low values of the kinetic coupling efficiencies are not consistent with theoretical predictions of both energy-conserving mechanisms \citep[0.5\% $-$ 5\%;][]{costat2014} and radiation-pressure driven scenario \citep[0.1\% $-$ 1\%;][]{costat2018}. 
Overall, the kinetic coupling efficiencies inferred from our results are an indication that the outflow observed is not very relevant from the energetic point of view. This study only considers ionised outflow and it could be that more massive outflows with high coupling efficiencies are being driven in the molecular or other phases. For example in \cite{fiore2017}, the molecular phase gives mass outflow rates that are higher than the ionised phase. This implies that the outflows in this quasar could be more relevant in other gas phases. 

 The momentum boosts obtained in this study are too low compared to predictions for either energy ($\sim$20) or momentum ($\sim$1) conserving mechanisms. With the uncertainties that arise from different assumptions in measuring outflow properties such as outflow luminosity, electron density, radius, and velocity, we can not draw firm conclusions about the driving mechanisms in this system. Observation from other wavebands is necessary to explore outflows of this source in other phases and the driving mechanisms for example radio and deeper X-rays. 

We measure ionised mass loading factor using, MLF$\;=\;\dot{\rm{M}}^{\rm{ion}}_{\rm{out}}/SFR$. We used the SFR, 2 M$_{\odot}$ yr$^{-1}$ from \cite{brusa2022}. This gives MLF of five which implies that the outflow plays a vital role in AGN feedback and can potentially affect the gas in the ISM faster than star formation and also an indication that the outflows observed are likely AGN-driven than SF-driven. Values of MLF that indicate such scenarios have been reported in the literature \citep[e.g.][]{venturi2023,speranza2023} for obscured quasars at low redshifts.

 The improved resolution and sensitivity of MUSE data allow for more accurate measurements of the relevant parameters such as radius, velocity and [O III] luminosity. Despite the several assumptions made, the spatially resolved data enables improved estimates of the mass outflow rate, energetics and insights into the environment around the quasar.

\section{Conclusions}
\label{conclusions}
In this study, we performed a detailed analysis of the morphology and kinematics of the [O III] emission line in object ID 608. The data used in this analysis were obtained through MUSE observations for program 108.22M6.001 period 108.
Our findings are summarised as follows.
\begin{itemize}[label=-]
    \item Our source is part of an interacting system featuring two close-by companion galaxies and a more distant galaxy situated within $\sim$50 projected kpc away from the nucleus of the quasar as identified for the first time in this study. These companions include C1 which shows signatures of ionisation from the quasar, C2, and C3 with a spectrum that resembles that of an elliptical galaxy. The observed extended emission emanating from our source in the SW can be attributed to a cloud of gas, which is highly probable to have originated from the ongoing gravitational interactions or merging transpiring within the system. 
    
    \item We have observed ionised outflows in our quasar with significant velocities extending up to  $\sim 8.7^{+0.4}_{-0.8}$ $-$ $9.2^{+1.2}_{-0.4}$ kpc. These outflows exhibit both blue- and red-shifted velocities, with the blue-shifted component reaching speeds as high as $-$1200\,$\; \rm{km\;s^{-1}}$ and the red-shifted component reaching speeds of up to 1000\,$\; \rm{km\;s^{-1}}$. For the first time, we have discovered the presence of a blueshifted counterpart to this outflow.
    
    \item Mass outflow rates were computed using the [O III] luminosity, assuming an electron density of 500 cm$^{-3}$, estimating the radius and velocity. The resulting ionised mass outflow rates were determined to be $10$ M$_{\odot}$ yr$^{-1}$, with a corresponding kinetic power of 2$\times$10$^{42}$\,$\;\rm{erg\;s^{-1}}$. The mass loading factor was estimated to be five which is an indication that these outflows are more likely to be AGN-driven than star formation-driven.
   
    \item  The kinetic coupling efficiencies range from $\sim$ 0.01\% $-$0.03\%. The momentum boosts are $\sim$ 0.2. These values are low, indicating that the ionised outflow is not very relevant from the energetic point of view. 
 
\end{itemize}

Our findings emphasise the importance of utilising integral field unit (IFU) and spatially resolved data to provide valuable insights into the morphology of outflows and better estimate the AGN outflow properties better. The morphology of the outflow is important to understand since it influences how they interact and couple with the ISM, whether it be radiatively driven or kinetic driven outflows. The interplay of interactions, mergers, and ionised gas within our system and the kinematics morphology renders this system interesting for further investigation in other wavelengths such as radio or deeper observation with Chandra to understand the nature of the companions. The discovery of this interactive nature around our quasar points to the importance of utilizing X-ray selection to probe outflows in red and obscured quasars which may be buried in such systems. 

\begin{acknowledgements}

BM, GS, IEL, BL, IMR and CA are supported by the European Union's Innovative Training  Network (ITN) funded by the  Marie Sklodowska-Curie Actions in Horizon 2020 No 860744 (BiD4BEST). MB acknowledges support from PRIN MIUR 2017PH3WAT ('Black hole winds and the baryon life cycle of galaxies').

CRA acknowledges support from the projects 'Feeding and
feedback in active galaxies', with reference PID2019-106027GB-C42,
funded by MICINN-AEI/10.13039/501100011033 and 'Quantifying the
impact of quasar feedback on galaxy evolution', with reference
EUR2020-112266, funded by MICINN-AEI/10.13039/501100011033 and the
European Union NextGenerationEU/PRTR.

MP acknowledges support from the research project PID2021- 127718NB-I00 of the Spanish Ministry of Science and Innovation or State Agency of Research (MCIN/AEI/10.13039/501100011033).

AL is partly supported by:
'Data Science methods for MultiMessenger Astrophysics \& Multi-Survey Cosmology',
Programmazione triennale 2021/2023 (DM n.2503 dd. 9 December 2019), Programma
Congiunto Scuole; Fondazione ICSC, Spoke 3 Astrophysics and Cosmos
Observations Project ID CN-00000013; INAF Large Grant 2022 funding scheme, project
MeerKAT and LOFAR Team up: a Unique Radio Window on Galaxy or AGN co-Evolution.
AL thanks M.V. Zanchettin for useful discussions.

\end{acknowledgements}

%
%
\bibliographystyle{aa}
\bibliography{example}

\begin{thebibliography}{78}
\expandafter\ifx\csname natexlab\endcsname\relax\def\natexlab#1{#1}\fi

\bibitem[{{Alexander} \& {Hickox}(2012)}]{alexanderDM2012}
{Alexander}, D.~M. \& {Hickox}, R.~C. 2012, \nar, 56, 93

\bibitem[{{Bacon} {et~al.}(2010){Bacon}, {Accardo}, {Adjali}, {Anwand}, {Bauer}, {Biswas}, {Blaizot}, {Boudon}, {Brau-Nogue}, {Brinchmann}, {Caillier}, {Capoani}, {Carollo}, {Contini}, {Couderc}, {Daguis{\'e}}, {Deiries}, {Delabre}, {Dreizler}, {Dubois}, {Dupieux}, {Dupuy}, {Emsellem}, {Fechner}, {Fleischmann}, {Fran{\c{c}}ois}, {Gallou}, {Gharsa}, {Glindemann}, {Gojak}, {Guiderdoni}, {Hansali}, {Hahn}, {Jarno}, {Kelz}, {Koehler}, {Kosmalski}, {Laurent}, {Le Floch}, {Lilly}, {Lizon}, {Loupias}, {Manescau}, {Monstein}, {Nicklas}, {Olaya}, {Pares}, {Pasquini}, {P{\'e}contal-Rousset}, {Pell{\'o}}, {Petit}, {Popow}, {Reiss}, {Remillieux}, {Renault}, {Roth}, {Rupprecht}, {Serre}, {Schaye}, {Soucail}, {Steinmetz}, {Streicher}, {Stuik}, {Valentin}, {Vernet}, {Weilbacher}, {Wisotzki}, \& {Yerle}}]{bacon2010}
{Bacon}, R., {Accardo}, M., {Adjali}, L., {et~al.} 2010, in Society of Photo-Optical Instrumentation Engineers (SPIE) Conference Series, Vol. 7735, Ground-based and Airborne Instrumentation for Astronomy III, ed. I.~S. {McLean}, S.~K. {Ramsay}, \& H.~{Takami}, 773508

\bibitem[{{Blecha} {et~al.}(2018){Blecha}, {Snyder}, {Satyapal}, \& {Ellison}}]{blecha2018}
{Blecha}, L., {Snyder}, G.~F., {Satyapal}, S., \& {Ellison}, S.~L. 2018, \mnras, 478, 3056

\bibitem[{{Brunner} {et~al.}(2022){Brunner}, {Liu}, {Lamer}, {Georgakakis}, {Merloni}, {Brusa}, {Bulbul}, {Dennerl}, {Friedrich}, {Liu}, {Maitra}, {Nandra}, {Ramos-Ceja}, {Sanders}, {Stewart}, {Boller}, {Buchner}, {Clerc}, {Comparat}, {Dwelly}, {Eckert}, {Finoguenov}, {Freyberg}, {Ghirardini}, {Gueguen}, {Haberl}, {Kreykenbohm}, {Krumpe}, {Osterhage}, {Pacaud}, {Predehl}, {Reiprich}, {Robrade}, {Salvato}, {Santangelo}, {Schrabback}, {Schwope}, \& {Wilms}}]{brunner2022}
{Brunner}, H., {Liu}, T., {Lamer}, G., {et~al.} 2022, \aap, 661, A1

\bibitem[{{Brusa} {et~al.}(2016){Brusa}, {Perna}, {Cresci}, {Schramm}, {Delvecchio}, {Lanzuisi}, {Mainieri}, {Mignoli}, {Zamorani}, {Berta}, {Bongiorno}, {Comastri}, {Fiore}, {Kakkad}, {Marconi}, {Rosario}, {Contini}, \& {Lamareille}}]{brusa2016}
{Brusa}, M., {Perna}, M., {Cresci}, G., {et~al.} 2016, \aap, 588, A58

\bibitem[{{Brusa} {et~al.}(2022){Brusa}, {Urrutia}, {Toba}, {Buchner}, {Li}, {Liu}, {Perna}, {Salvato}, {Merloni}, {Musiimenta}, {Nandra}, {Wolf}, {Arcodia}, {Dwelly}, {Georgakakis}, {Goulding}, {Matsuoka}, {Nagao}, {Schramm}, {Silverman}, \& {Terashima}}]{brusa2022}
{Brusa}, M., {Urrutia}, T., {Toba}, Y., {et~al.} 2022, \aap, 661, A9

\bibitem[{{Calzetti} {et~al.}(2000){Calzetti}, {Armus}, {Bohlin}, {Kinney}, {Koornneef}, \& {Storchi-Bergmann}}]{calzetti2000}
{Calzetti}, D., {Armus}, L., {Bohlin}, R.~C., {et~al.} 2000, \apj, 533, 682

\bibitem[{{Cano-D{\'\i}az} {et~al.}(2012){Cano-D{\'\i}az}, {Maiolino}, {Marconi}, {Netzer}, {Shemmer}, \& {Cresci}}]{canodiaz2012}
{Cano-D{\'\i}az}, M., {Maiolino}, R., {Marconi}, A., {et~al.} 2012, \aap, 537, L8

\bibitem[{{Cappellari} \& {Copin}(2003)}]{cappellari2003}
{Cappellari}, M. \& {Copin}, Y. 2003, \mnras, 342, 345

\bibitem[{{Cardelli} {et~al.}(1989){Cardelli}, {Clayton}, \& {Mathis}}]{cardelli1989}
{Cardelli}, J.~A., {Clayton}, G.~C., \& {Mathis}, J.~S. 1989, \apj, 345, 245

\bibitem[{{Carniani} {et~al.}(2013){Carniani}, {Marconi}, {Biggs}, {Cresci}, {Cupani}, {D'Odorico}, {Humphreys}, {Maiolino}, {Mannucci}, {Molaro}, {Nagao}, {Testi}, \& {Zwaan}}]{carniani2013}
{Carniani}, S., {Marconi}, A., {Biggs}, A., {et~al.} 2013, \aap, 559, A29

\bibitem[{{Carniani} {et~al.}(2015){Carniani}, {Marconi}, {Maiolino}, {Balmaverde}, {Brusa}, {Cano-D{\'\i}az}, {Cicone}, {Comastri}, {Cresci}, {Fiore}, {Feruglio}, {La Franca}, {Mainieri}, {Mannucci}, {Nagao}, {Netzer}, {Piconcelli}, {Risaliti}, {Schneider}, \& {Shemmer}}]{carniani2015}
{Carniani}, S., {Marconi}, A., {Maiolino}, R., {et~al.} 2015, \aap, 580, A102

\bibitem[{{Cicone} {et~al.}(2018){Cicone}, {Brusa}, {Ramos Almeida}, {Cresci}, {Husemann}, \& {Mainieri}}]{Cicone2018}
{Cicone}, C., {Brusa}, M., {Ramos Almeida}, C., {et~al.} 2018, Nature Astronomy, 2, 176

\bibitem[{{Costa} {et~al.}(2018){Costa}, {Rosdahl}, {Sijacki}, \& {Haehnelt}}]{costat2018}
{Costa}, T., {Rosdahl}, J., {Sijacki}, D., \& {Haehnelt}, M.~G. 2018, \mnras, 479, 2079

\bibitem[{{Costa} {et~al.}(2014){Costa}, {Sijacki}, \& {Haehnelt}}]{costat2014}
{Costa}, T., {Sijacki}, D., \& {Haehnelt}, M.~G. 2014, \mnras, 444, 2355

\bibitem[{{Cresci} {et~al.}(2015){Cresci}, {Mainieri}, {Brusa}, {Marconi}, {Perna}, {Mannucci}, {Piconcelli}, {Maiolino}, {Feruglio}, {Fiore}, {Bongiorno}, {Lanzuisi}, {Merloni}, {Schramm}, {Silverman}, \& {Civano}}]{cresci2015}
{Cresci}, G., {Mainieri}, V., {Brusa}, M., {et~al.} 2015, \apj, 799, 82

\bibitem[{{Cresci} {et~al.}(2023){Cresci}, {Tozzi}, {Perna}, {Brusa}, {Marconcini}, {Marconi}, {Carniani}, {Brienza}, {Giroletti}, {Belfiore}, {Ginolfi}, {Mannucci}, {Ulivi}, {Scholtz}, {Venturi}, {Arribas}, {{\"U}bler}, {D'Eugenio}, {Mingozzi}, {Balmaverde}, {Capetti}, {Parlanti}, \& {Zana}}]{cresci2023}
{Cresci}, G., {Tozzi}, G., {Perna}, M., {et~al.} 2023, \aap, 672, A128

\bibitem[{{Davies} {et~al.}(2020){Davies}, {Baron}, {Shimizu}, {Netzer}, {Burtscher}, {de Zeeuw}, {Genzel}, {Hicks}, {Koss}, {Lin}, {Lutz}, {Maciejewski}, {M{\"u}ller-S{\'a}nchez}, {Orban de Xivry}, {Ricci}, {Riffel}, {Riffel}, {Rosario}, {Schartmann}, {Schnorr-M{\"u}ller}, {Shangguan}, {Sternberg}, {Sturm}, {Storchi-Bergmann}, {Tacconi}, \& {Veilleux}}]{davies2020}
{Davies}, R., {Baron}, D., {Shimizu}, T., {et~al.} 2020, \mnras, 498, 4150

\bibitem[{{DeGraf} {et~al.}(2015){DeGraf}, {Di Matteo}, {Treu}, {Feng}, {Woo}, \& {Park}}]{degraf2015}
{DeGraf}, C., {Di Matteo}, T., {Treu}, T., {et~al.} 2015, \mnras, 454, 913

\bibitem[{{D'Eugenio} {et~al.}(2023){D'Eugenio}, {Perez-Gonzalez}, {Maiolino}, {Scholtz}, {Perna}, {Circosta}, {Uebler}, {Arribas}, {Boeker}, {Bunker}, {Carniani}, {Charlot}, {Chevallard}, {Cresci}, {Curtis-Lake}, {Jones}, {Kumari}, {Lamperti}, {Looser}, {Parlanti}, {Rix}, {Robertson}, {Rodriguez Del Pino}, {Tacchella}, {Venturi}, \& {Willott}}]{francesco2023}
{D'Eugenio}, F., {Perez-Gonzalez}, P., {Maiolino}, R., {et~al.} 2023, arXiv e-prints, arXiv:2308.06317

\bibitem[{{Di Matteo} {et~al.}(2005){Di Matteo}, {Springel}, \& {Hernquist}}]{dimatteo2005}
{Di Matteo}, T., {Springel}, V., \& {Hernquist}, L. 2005, \nat, 433, 604

\bibitem[{{Duncan} {et~al.}(2023){Duncan}, {Windhorst}, {Koekemoer}, {R{\"o}ttgering}, {Cohen}, {Jansen}, {Summers}, {Tompkins}, {Hutchison}, {Conselice}, {Driver}, {Yan}, {Adams}, {Cheng}, {Coe}, {Diego}, {Dole}, {Frye}, {Gim}, {Grogin}, {Holwerda}, {Lim}, {Marshall}, {Nonino}, {Pirzkal}, {Robotham}, {Ryan}, \& {Willmer}}]{kenneth2023}
{Duncan}, K.~J., {Windhorst}, R.~A., {Koekemoer}, A.~M., {et~al.} 2023, \mnras, 522, 4548

\bibitem[{{Fawcett} {et~al.}(2020){Fawcett}, {Alexander}, {Rosario}, {Klindt}, {Fotopoulou}, {Lusso}, {Morabito}, \& {Calistro Rivera}}]{fawcett2020}
{Fawcett}, V.~A., {Alexander}, D.~M., {Rosario}, D.~J., {et~al.} 2020, \mnras, 494, 4802

\bibitem[{Fetherolf {et~al.}(2020)Fetherolf, Reddy, Shapley, Kriek, Siana, Coil, Mobasher, Freeman, Sanders, Price, Shivaei, Azadi, de Groot, Leung, \& Zick}]{tara2020}
Fetherolf, T., Reddy, N.~A., Shapley, A.~E., {et~al.} 2020, Monthly Notices of the Royal Astronomical Society, 498, 5009

\bibitem[{{Fiore} {et~al.}(2017){Fiore}, {Feruglio}, {Shankar}, {Bischetti}, {Bongiorno}, {Brusa}, {Carniani}, {Cicone}, {Duras}, {Lamastra}, {Mainieri}, {Marconi}, {Menci}, {Maiolino}, {Piconcelli}, {Vietri}, \& {Zappacosta}}]{fiore2017}
{Fiore}, F., {Feruglio}, C., {Shankar}, F., {et~al.} 2017, \aap, 601, A143

\bibitem[{{Freudling} {et~al.}(2013){Freudling}, {Romaniello}, {Bramich}, {Ballester}, {Forchi}, {Garc{\'\i}a-Dabl{\'o}}, {Moehler}, \& {Neeser}}]{freudling2013}
{Freudling}, W., {Romaniello}, M., {Bramich}, D.~M., {et~al.} 2013, \aap, 559, A96

\bibitem[{{Girdhar} {et~al.}(2022){Girdhar}, {Harrison}, {Mainieri}, {Bittner}, {Costa}, {Kharb}, {Mukherjee}, {Arrigoni Battaia}, {Alexander}, {Calistro Rivera}, {Circosta}, {De Breuck}, {Edge}, {Farina}, {Kakkad}, {Lansbury}, {Molyneux}, {Mullaney}, {S}, {Thomson}, \& {Ward}}]{girdhar2022}
{Girdhar}, A., {Harrison}, C.~M., {Mainieri}, V., {et~al.} 2022, \mnras, 512, 1608

\bibitem[{{Harrison}(2017)}]{harrison2017}
{Harrison}, C.~M. 2017, Nature Astronomy, 1, 0165

\bibitem[{{Harrison} \& {Ramos Almeida}(2024)}]{harrisonCM2024}
{Harrison}, C.~M. \& {Ramos Almeida}, C. 2024, arXiv e-prints, arXiv:2404.08050

\bibitem[{{Hervella Seoane} {et~al.}(2023){Hervella Seoane}, {Ramos Almeida}, {Acosta Pulido}, {Speranza}, {Tadhunter}, \& {Bessiere}}]{hervella2023}
{Hervella Seoane}, K., {Ramos Almeida}, C., {Acosta Pulido}, J.~A., {et~al.} 2023, arXiv e-prints, arXiv:2309.10572

\bibitem[{Hopkins \& Elvis(2009)}]{hopkins2009}
Hopkins, P.~F. \& Elvis, M. 2009, Monthly Notices of the Royal Astronomical Society, 401, 7

\bibitem[{{Hopkins} {et~al.}(2005){Hopkins}, {Hernquist}, {Cox}, {Di Matteo}, {Martini}, {Robertson}, \& {Springel}}]{hopkins2005}
{Hopkins}, P.~F., {Hernquist}, L., {Cox}, T.~J., {et~al.} 2005, \apj, 630, 705

\bibitem[{{Hopkins} {et~al.}(2008){Hopkins}, {Hernquist}, {Cox}, \& {Kere{\v{s}}}}]{hopkins2008}
{Hopkins}, P.~F., {Hernquist}, L., {Cox}, T.~J., \& {Kere{\v{s}}}, D. 2008, \apjs, 175, 356

\bibitem[{{Husemann} {et~al.}(2016){Husemann}, {Scharw{\"a}chter}, {Bennert}, {Mainieri}, {Woo}, \& {Kakkad}}]{husemann2016}
{Husemann}, B., {Scharw{\"a}chter}, J., {Bennert}, V.~N., {et~al.} 2016, \aap, 594, A44

\bibitem[{{Husemann} {et~al.}(2013){Husemann}, {Wisotzki}, {S{\'a}nchez}, \& {Jahnke}}]{Husemann2013}
{Husemann}, B., {Wisotzki}, L., {S{\'a}nchez}, S.~F., \& {Jahnke}, K. 2013, \aap, 549, A43

\bibitem[{{Kakkad} {et~al.}(2020){Kakkad}, {Mainieri}, {Vietri}, {Carniani}, {Harrison}, {Perna}, {Scholtz}, {Circosta}, {Cresci}, {Husemann}, {Bischetti}, {Feruglio}, {Fiore}, {Marconi}, {Padovani}, {Brusa}, {Cicone}, {Comastri}, {Lanzuisi}, {Mannucci}, {Menci}, {Netzer}, {Piconcelli}, {Puglisi}, {Salvato}, {Schramm}, {Silverman}, {Vignali}, {Zamorani}, \& {Zappacosta}}]{kakkad2020a}
{Kakkad}, D., {Mainieri}, V., {Vietri}, G., {et~al.} 2020, \aap, 642, A147

\bibitem[{{Kakkad} {et~al.}(2023){Kakkad}, {Mainieri}, {Vietri}, {Lamperti}, {Carniani}, {Cresci}, {Harrison}, {Marconi}, {Bischetti}, {Cicone}, {Circosta}, {Husemann}, {Man}, {Mannucci}, {Netzer}, {Padovani}, {Perna}, {Puglisi}, {Scholtz}, {Tozzi}, {Vignali}, \& {Zappacosta}}]{kakkad2023}
{Kakkad}, D., {Mainieri}, V., {Vietri}, G., {et~al.} 2023, \mnras, 520, 5783

\bibitem[{{Kakkad} {et~al.}(2022){Kakkad}, {Sani}, {Rojas}, {Mallmann}, {Veilleux}, {Bauer}, {Ricci}, {Mushotzky}, {Koss}, {Ricci}, {Treister}, {Privon}, {Nguyen}, {B{\"a}r}, {Harrison}, {Oh}, {Powell}, {Riffel}, {Stern}, {Trakhtenbrot}, \& {Urry}}]{kakkad2022}
{Kakkad}, D., {Sani}, E., {Rojas}, A.~F., {et~al.} 2022, \mnras, 511, 2105

\bibitem[{{Kang} \& {Woo}(2018)}]{kangandwoo2018}
{Kang}, D. \& {Woo}, J.-H. 2018, \apj, 864, 124

\bibitem[{{Kim} {et~al.}(2015){Kim}, {Im}, {Glikman}, {Woo}, \& {Urrutia}}]{kim2015}
{Kim}, D., {Im}, M., {Glikman}, E., {Woo}, J.-H., \& {Urrutia}, T. 2015, \apj, 812, 66

\bibitem[{{King} \& {Pounds}(2015)}]{kingp2015}
{King}, A. \& {Pounds}, K. 2015, \araa, 53, 115

\bibitem[{{Klindt} {et~al.}(2019){Klindt}, {Alexander}, {Rosario}, {Lusso}, \& {Fotopoulou}}]{klindt2019}
{Klindt}, L., {Alexander}, D.~M., {Rosario}, D.~J., {Lusso}, E., \& {Fotopoulou}, S. 2019, \mnras, 488, 3109

\bibitem[{{Kolcu} {et~al.}(2023){Kolcu}, {Maciejewski}, {Gadotti}, {Fragkoudi}, {Erwin}, {S{\'a}nchez-Bl{\'a}zquez}, {Neumann}, {Van de Ven}, {de S{\'a}-Freitas}, {Longmore}, \& {Debattista}}]{kolcu2023}
{Kolcu}, T., {Maciejewski}, W., {Gadotti}, D.~A., {et~al.} 2023, \mnras, 524, 207

\bibitem[{{Kormendy} \& {Ho}(2013)}]{kormendy2013}
{Kormendy}, J. \& {Ho}, L.~C. 2013, \araa, 51, 511

\bibitem[{{Lynden-Bell}(1969)}]{lyndenbell1969}
{Lynden-Bell}, D. 1969, \nat, 223, 690

\bibitem[{{McConnell} \& {Ma}(2013)}]{mcconnell2013}
{McConnell}, N.~J. \& {Ma}, C.-P. 2013, \apj, 764, 184

\bibitem[{{Merloni} {et~al.}(2024){Merloni}, {Lamer}, {Ramos-Ceja}, {Brunner}, {Buchner}, \& {et}}]{merloni2024}
{Merloni}, A., {Lamer}, G., {Ramos-Ceja}, M., {et~al.} 2024, \aap, 682, A34

\bibitem[{{Mullaney} {et~al.}(2013){Mullaney}, {Alexander}, {Fine}, {Goulding}, {Harrison}, \& {Hickox}}]{mullaney2013}
{Mullaney}, J.~R., {Alexander}, D.~M., {Fine}, S., {et~al.} 2013, \mnras, 433, 622

\bibitem[{{Musiimenta} {et~al.}(2023){Musiimenta}, {Brusa}, {Liu}, {Salvato}, {Buchner}, {Igo}, {Waddell}, {Toba}, {Arcodia}, {Comparat}, {Alexander}, {Shankar}, {Lapi}, {Ramos Almeida}, {Georgakakis}, {Merloni}, {Urrutia}, {Li}, {Terashima}, {Shen}, {Wu}, {Dwelly}, {Nandra}, \& {Wolf}}]{musiimenta2023}
{Musiimenta}, B., {Brusa}, M., {Liu}, T., {et~al.} 2023, \aap, 679, A84

\bibitem[{{Nandra} {et~al.}(2024){Nandra}, {Waddell}, {Liu}, {Buchner}, {Dwelly}, {Salvato}, {Shen}, {Wu}, {Arcodia}, {Boller}, {Brunner}, {Brusa}, {Collmar}, {Comparat}, {Georgakakis}, {Grau}, {H{\"a}mmerich}, {Ibarra-Medel}, {Igo}, {Krumpe}, {Lamer}, {Merloni}, {Musiimenta}, {Wolf}, {Assef}, {Bauer}, {Brandt}, \& {Rix}}]{nandra2024}
{Nandra}, K., {Waddell}, S.~G.~H., {Liu}, T., {et~al.} 2024, arXiv e-prints, arXiv:2401.17300

\bibitem[{{Olivares} {et~al.}(2022){Olivares}, {Salom{\'e}}, {Hamer}, {Combes}, {Gaspari}, {Kolokythas}, {O'Sullivan}, {Beckmann}, {Babul}, {Polles}, {Lehnert}, {Loubser}, {Donahue}, {Gendron-Marsolais}, {Lagos}, {Pineau des Forets}, {Godard}, {Rose}, {Tremblay}, {Ferland}, \& {Guillard}}]{olivares2022}
{Olivares}, V., {Salom{\'e}}, P., {Hamer}, S.~L., {et~al.} 2022, \aap, 666, A94

\bibitem[{{Osterbrock}(1981)}]{osterbrock1981}
{Osterbrock}, D.~E. 1981, \apj, 249, 462

\bibitem[{{Ott}(2012)}]{ott2012}
{Ott}, T. 2012, {QFitsView: FITS file viewer}, Astrophysics Source Code Library, record ascl:1210.019

\bibitem[{{Perna} {et~al.}(2023){Perna}, {Arribas}, {Lamperti}, {Circosta}, {Bertola}, {P{\'e}rez-Gonz{\'a}lez}, {D'Eugenio}, {{\"U}bler}, {Cresci}, {Maiolino}, {Rodr{\'\i}guez Del Pino}, {Bunker}, {Charlot}, {Willott}, {Carniani}, {B{\"o}ker}, {Chevallard}, {Curti}, {Jones}, {Kumari}, {Marshall}, {Saxena}, {Scholtz}, {Venturi}, \& {Witstok}}]{Michele2023b}
{Perna}, M., {Arribas}, S., {Lamperti}, I., {et~al.} 2023, arXiv e-prints, arXiv:2310.03067

\bibitem[{{Perna} {et~al.}(2015){Perna}, {Brusa}, {Salvato}, {Cresci}, {Lanzuisi}, {Berta}, {Delvecchio}, {Fiore}, {Lutz}, {Le Floc'h}, {Mainieri}, \& {Riguccini}}]{perna2015}
{Perna}, M., {Brusa}, M., {Salvato}, M., {et~al.} 2015, \aap, 583, A72

\bibitem[{{Perna} {et~al.}(2017){Perna}, {Lanzuisi}, {Brusa}, {Cresci}, \& {Mignoli}}]{perna2017b}
{Perna}, M., {Lanzuisi}, G., {Brusa}, M., {Cresci}, G., \& {Mignoli}, M. 2017, \aap, 606, A96

\bibitem[{{Predehl} {et~al.}(2021){Predehl}, {Andritschke}, {Arefiev}, {Babyshkin}, {Batanov}, {Becker}, {B{\"o}hringer}, {Bogomolov}, {Boller}, {Borm}, {Bornemann}, {Br{\"a}uninger}, {Br{\"u}ggen}, {Brunner}, {Brusa}, {Bulbul}, {Buntov}, {Burwitz}, {Burkert}, {Clerc}, {Churazov}, {Coutinho}, {Dauser}, {Dennerl}, {Doroshenko}, {Eder}, {Emberger}, {Eraerds}, {Finoguenov}, {Freyberg}, {Friedrich}, {Friedrich}, {F{\"u}rmetz}, {Georgakakis}, {Gilfanov}, {Granato}, {Grossberger}, {Gueguen}, {Gureev}, {Haberl}, {H{\"a}lker}, {Hartner}, {Hasinger}, {Huber}, {Ji}, {Kienlin}, {Kink}, {Korotkov}, {Kreykenbohm}, {Lamer}, {Lomakin}, {Lapshov}, {Liu}, {Maitra}, {Meidinger}, {Menz}, {Merloni}, {Mernik}, {Mican}, {Mohr}, {M{\"u}ller}, {Nandra}, {Nazarov}, {Pacaud}, {Pavlinsky}, {Perinati}, {Pfeffermann}, {Pietschner}, {Ramos-Ceja}, {Rau}, {Reiffers}, {Reiprich}, {Robrade}, {Salvato}, {Sanders}, {Santangelo}, {Sasaki}, {Scheuerle}, {Schmid}, {Schmitt}, {Schwope}, {Shirshakov}, {Steinmetz}, {Stewart}, {Str{\"u}der},
  {Sunyaev}, {Tenzer}, {Tiedemann}, {Tr{\"u}mper}, {Voron}, {Weber}, {Wilms}, \& {Yaroshenko}}]{predehl2021}
{Predehl}, P., {Andritschke}, R., {Arefiev}, V., {et~al.} 2021, \aap, 647, A1

\bibitem[{{Rees}(1984)}]{rees1984}
{Rees}, M.~J. 1984, \araa, 22, 471

\bibitem[{{Reyes} {et~al.}(2008){Reyes}, {Zakamska}, {Strauss}, {Green}, {Krolik}, {Shen}, {Richards}, {Anderson}, \& {Schneider}}]{reyes2008}
{Reyes}, R., {Zakamska}, N.~L., {Strauss}, M.~A., {et~al.} 2008, \aj, 136, 2373

\bibitem[{{Salvato} {et~al.}(2022){Salvato}, {Wolf}, {Dwelly}, {Georgakakis}, {Brusa}, {Merloni}, {Liu}, {Toba}, {Nandra}, {Lamer}, {Buchner}, {Schneider}, {Freund}, {Rau}, {Schwope}, {Nishizawa}, {Klein}, {Arcodia}, {Comparat}, {Musiimenta}, {Nagao}, {Brunner}, {Malyali}, {Finoguenov}, {Anderson}, {Shen}, {Ibarra-Mendel}, {Trump}, {Brandt}, {Urry}, {Rivera}, {Krumpe}, {Urrutia}, {Miyaji}, {Ichikawa}, {Schneider}, {Fresco}, {Wilms}, {Boller}, {Haase}, {Brownstein}, {Lane}, {Bizyaev}, \& {Nitschelm}}]{salvato2021}
{Salvato}, M., {Wolf}, J., {Dwelly}, T., {et~al.} 2022, arXiv e-prints, arXiv:2106.14520

\bibitem[{{Scholtz} {et~al.}(2020){Scholtz}, {Harrison}, {Rosario}, {Alexander}, {Chen}, {Kakkad}, {Mainieri}, {Tiley}, {Turner}, {Cirasuolo}, {Sharples}, \& {Stach}}]{scholtz2020}
{Scholtz}, J., {Harrison}, C.~M., {Rosario}, D.~J., {et~al.} 2020, \mnras, 492, 3194

\bibitem[{{Shen} {et~al.}(2023){Shen}, {Liu}, {He}, {Zakamska}, {Glikman}, {Greene}, {Hu}, {Mou}, {Wylezalek}, \& {Rupke}}]{shen2023}
{Shen}, L., {Liu}, G., {He}, Z., {et~al.} 2023, arXiv e-prints, arXiv:2307.06059

\bibitem[{{Silk} \& {Rees}(1998)}]{silk1998}
{Silk}, J. \& {Rees}, M.~J. 1998, \aap, 331, L1

\bibitem[{{Speranza} {et~al.}(2024){Speranza}, {Ramos Almeida}, {Acosta-Pulido}, {Audibert}, {Holden}, {Tadhunter}, {Lapi}, {Gonz{\'a}lez-Mart{\'\i}n}, {Brusa}, {L{\'o}pez}, {Musiimenta}, \& {Shankar}}]{speranza2023}
{Speranza}, G., {Ramos Almeida}, C., {Acosta-Pulido}, J.~A., {et~al.} 2024, \aap, 681, A63

\bibitem[{{Speranza} {et~al.}(2022){Speranza}, {Ramos Almeida}, {Acosta-Pulido}, {Riffel}, {Tadhunter}, {Pierce}, {Rodr{\'\i}guez-Ardila}, {Coloma Puga}, {Brusa}, {Musiimenta}, {Alexander}, {Lapi}, {Shankar}, \& {Villforth}}]{speranza2022}
{Speranza}, G., {Ramos Almeida}, C., {Acosta-Pulido}, J.~A., {et~al.} 2022, \aap, 665, A55

\bibitem[{{Spergel} {et~al.}(2003){Spergel}, {Verde}, {Peiris}, {Komatsu}, {Nolta}, {Bennett}, {Halpern}, {Hinshaw}, {Jarosik}, {Kogut}, {Limon}, {Meyer}, {Page}, {Tucker}, {Weiland}, {Wollack}, \& {Wright}}]{spergel2003}
{Spergel}, D.~N., {Verde}, L., {Peiris}, H.~V., {et~al.} 2003, \apjs, 148, 175

\bibitem[{{Toba} {et~al.}(2017){Toba}, {Bae}, {Nagao}, {Woo}, {Wang}, {Wagner}, {Sun}, \& {Chang}}]{toba2017}
{Toba}, Y., {Bae}, H.-J., {Nagao}, T., {et~al.} 2017, \apj, 850, 140

\bibitem[{{Toba} {et~al.}(2022){Toba}, {Yamada}, {Matsubayashi}, {Terao}, {Moriya}, {Ueda}, {Ohta}, {Hashiguchi}, {Himoto}, {Izumiura}, {Joh}, {Kato}, {Koyama}, {Maehara}, {Misato}, {Noboriguchi}, {Ogawa}, {Ota}, {Shibata}, {Tamada}, {Yanagawa}, {Yonekura}, {Nagao}, {Akiyama}, {Kajisawa}, \& {Matsuoka}}]{toba2022}
{Toba}, Y., {Yamada}, S., {Matsubayashi}, K., {et~al.} 2022, \pasj, 74, 1356

\bibitem[{{Tremaine} {et~al.}(2002){Tremaine}, {Gebhardt}, {Bender}, {Bower}, {Dressler}, {Faber}, {Filippenko}, {Green}, {Grillmair}, {Ho}, {Kormendy}, {Lauer}, {Magorrian}, {Pinkney}, \& {Richstone}}]{tremaine2002}
{Tremaine}, S., {Gebhardt}, K., {Bender}, R., {et~al.} 2002, \apj, 574, 740

\bibitem[{{Urrutia} {et~al.}(2012){Urrutia}, {Lacy}, {Spoon}, {Glikman}, {Petric}, \& {Schulz}}]{urrutia2012}
{Urrutia}, T., {Lacy}, M., {Spoon}, H., {et~al.} 2012, \apj, 757, 125

\bibitem[{{Venturi} {et~al.}(2023){Venturi}, {Treister}, {Finlez}, {D'Ago}, {Bauer}, {Harrison}, {Ramos Almeida}, {Revalski}, {Ricci}, {Sartori}, {Girdhar}, {Keel}, \& {Tub{\'\i}n}}]{venturi2023}
{Venturi}, G., {Treister}, E., {Finlez}, C., {et~al.} 2023, \aap, 678, A127

\bibitem[{{Waddell} {et~al.}(2023){Waddell}, {Nandra}, {Buchner}, {Wu}, {Shen}, {Arcodia}, {Merloni}, {Salvato}, {Dauser}, {Boller}, {Liu}, {Comparat}, {Wolf}, {Dwelly}, {Ricci}, {Brownstein}, \& {Brusa}}]{Waddell2023}
{Waddell}, S. G.~H., {Nandra}, K., {Buchner}, J., {et~al.} 2023, arXiv e-prints, arXiv:2306.00961

\bibitem[{{Weilbacher} {et~al.}(2020){Weilbacher}, {Palsa}, {Streicher}, {Bacon}, {Urrutia}, {Wisotzki}, {Conseil}, {Husemann}, {Jarno}, {Kelz}, {P{\'e}contal-Rousset}, {Richard}, {Roth}, {Selman}, \& {Vernet}}]{weilbacher2020}
{Weilbacher}, P.~M., {Palsa}, R., {Streicher}, O., {et~al.} 2020, \aap, 641, A28

\bibitem[{Wylezalek {et~al.}(2022)Wylezalek, Vayner, Rupke, Zakamska, Veilleux, Ishikawa, Bertemes, Liu, Barrera-Ballesteros, Chen, Goulding, Greene, Hainline, Hamann, Heckman, Johnson, Lutz, Lützgendorf, Mainieri, Maiolino, Nesvadba, Ogle, \& Sturm}]{Wylezalek_2022}
Wylezalek, D., Vayner, A., Rupke, D. S.~N., {et~al.} 2022, The Astrophysical Journal Letters, 940, L7

\bibitem[{Zakamska {et~al.}(2016)Zakamska, Hamann, Pâris, Brandt, Greene, Strauss, Villforth, Wylezalek, Alexandroff, \& Ross}]{zakamska2016}
Zakamska, N.~L., Hamann, F., Pâris, I., {et~al.} 2016, Monthly Notices of the Royal Astronomical Society, 459, 3144

\bibitem[{{Zanchettin} {et~al.}(2023){Zanchettin}, {Feruglio}, {Massardi}, {Lapi}, {Bischetti}, {Cantalupo}, {Fiore}, {Bongiorno}, {Malizia}, {Marinucci}, {Molina}, {Piconcelli}, {Tombesi}, {Travascio}, {Tozzi}, \& {Tripodi}}]{zanchettin2023}
{Zanchettin}, M.~V., {Feruglio}, C., {Massardi}, M., {et~al.} 2023, \aap, 679, A88

\bibitem[{Zheng {et~al.}(2023)Zheng, Shi, Bian, Yu, Wang, Chen, Li, \& Gu}]{zheng2023}
Zheng, Z., Shi, Y., Bian, F., {et~al.} 2023, Monthly Notices of the Royal Astronomical Society, 523, 3274

\bibitem[{{Zhu} {et~al.}(2023){Zhu}, {Li}, {Cao}, {Lu}, {Cappellari}, \& {Mao}}]{zhu2023}
{Zhu}, K., {Li}, R., {Cao}, X., {et~al.} 2023, Research in Astronomy and Astrophysics, 23, 085001

\end{thebibliography}

\begin{appendix}{} 
\section{Velocity channels considering 100\,$\; \rm{km\;s^{-1}}$ bin size.  }
\label{Velocity_channels}

\begin{figure}[h!]
\centering
    \includegraphics[width=0.99\textwidth]{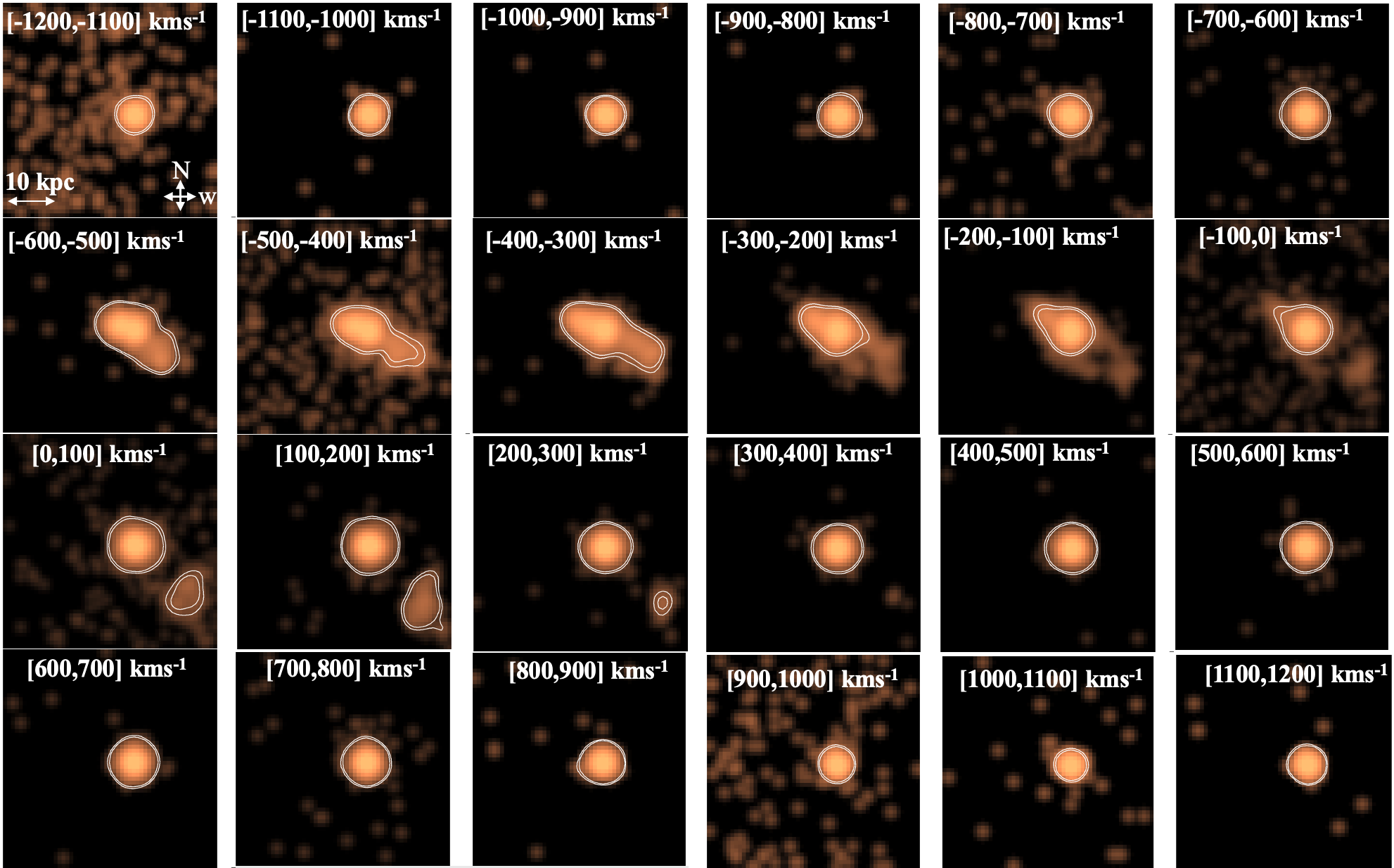} 
\caption[Velocity channels considering smaller bins of 100\,$\; \rm{km\;s^{-1}}$.]{ Velocity channels considering smaller bins of 100\,$\; \rm{km\;s^{-1}}$ on logarithmic scale. The white contours indicate the 2$\sigma$ and 3$\sigma$. 
}
\label{manyvelocitymaps}
\end{figure}

The extracted flux maps from various velocity bins of 100\,$\; \rm{km\;s^{-1}}$ from $-$1200\,$\; \rm{km\;s^{-1}}$ to 1200\,$\; \rm{km\;s^{-1}}$ are shown in Fig. \ref{manyvelocitymaps}. In this Figure, the different structures towards the NE and SW directions are visible from bins [$-$600, $-$500] $\rm{km\;s^{-1}}$ to [200, 300] $\rm{km\;s^{-1}}$. The rest of the bins have the [O III] emission not extending far away from the nucleus of the quasar.

\FloatBarrier

\section{Flux maps and velocity maps from three-Gaussian fitting and Voronoi binning  }
\label{kinematicsAP}

\subsection{Three-Gaussian fitting without Voronoi binning}
\label{multiGaussian}

\begin{figure*}[h!]
\centering
    \includegraphics[width=\textwidth]{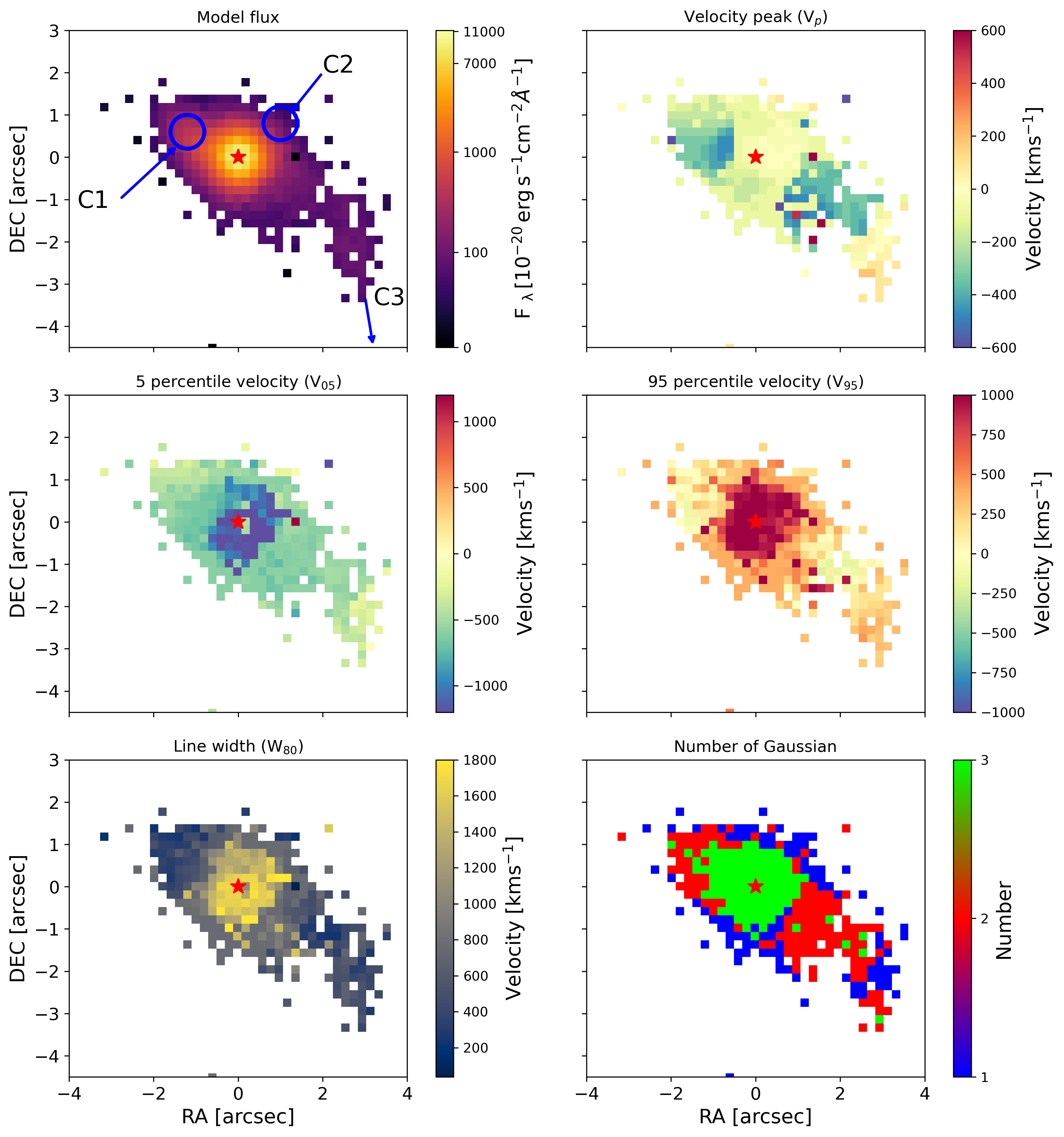} 
\caption[The results from the spaxel-by-spaxel three-Gaussian fitting.]{Results from the spaxel-by-spaxel three-Gaussian fitting of $8'' \times 7.5'' \cong$ 53 kpc $\times$ 50 kpc region around the quasar. Top left panel: the flux map. Top right panel: the velocity peak. In the middle panel, we present the V$_{\rm{05}}$ and V$_{\rm{95}}$ maps indicating the velocities at the 5$^{\rm{th}}$ and 95$^{\rm{th}}$ percentile, respectively. Bottom left panel: the line width (W$_{\rm{80}}$). Bottom right panel: the number of Gaussians fitted in each pixel. }
\label{velocitymaps3g}
\end{figure*}

 We performed spaxel-by-spaxel fitting with a maximum of three Gaussian components (n=1,2,3) around [O III] doublet ([O III]$\lambda$5007 and [O III]$\lambda$4949) using the \textit{lmfit} Python package and the in-house script described in \cite{speranza2023}. The three-Gaussian components are tuned in a such way as to avoid underfitting in the central regions and overfitting in the outer regions of the quasar. In this method, more than one Gaussian is fitted when the $\chi^{2}$ improves more than 10\% with respect to the previous fit. It is worth noting that the parameters of both emission lines were tied together and the flux ratios between the $\mathrm{[O III]\lambda{4959}}$ and $\mathrm{[O III]\lambda{5007}}$ lines were fixed at a ratio of 1:3 \citep{osterbrock1981}. We set two narrow components with velocity dispersions ($\sigma$) less than 500\,$\; \rm{km\;s^{-1}}$ and one broader component with $\sigma$ greater than 500\,$\; \rm{km\;s^{-1}}$. The velocity dispersions of each kinematic component can vary within these specified limits. The resulting flux map and kinematic maps are shown in Fig.$\;$\ref{velocitymaps3g}. These results are similar to the ones presented in Sect. \ref{kinematics_results}.

\FloatBarrier

\subsection{Voronoi bins data fitting with three-Gaussian}
\label{voronoi}

Another approach to visualising the kinematics and morphology of outflowing gas in AGNs is by grouping spaxels into bins. A commonly used method for this is adaptive spatial binning using Voronoi tessellations, which was introduced by \cite{cappellari2003}. The goal of this method is to achieve a constant signal-to-noise ratio (S/N) per bin, which is crucial for accurately analysing IFU data. Compared to analysing spaxel-by-spaxel, this approach allows for more efficient and reliable extraction of kinematic information. Analysing each spaxel separately can be computationally intensive and can produce noisy and biased results if the S/N is not sufficiently high. In contrast, adaptive spatial binning provides a constant S/N per bin, minimising the variance of the S/N ratio within each bin and enabling more reliable and efficient extraction of kinematics.

To implement this method, the desired target S/N for each bin and the minimum S/N required for a spaxel to be included in a bin must be set. Previous studies have employed this approach with various target S/N values, such as 10, 12, 15, 30, 40, and 50, for different analyses, including gas kinematics \citep[e.g.][]{girdhar2022,speranza2022,olivares2022,francesco2023,zheng2023,kolcu2023}, stellar populations \citep[][and referenced therein]{tara2020}, and resolved optical properties of high-redshift galaxies and others \citep{kenneth2023,zhu2023}. We have utilised this approach on our data cube to enhance the S/N. We binned our data cubes to achieve an average S/N of five per wavelength channel in each bin. Spaxels with S/N below three are discarded. We chose a relatively low threshold compared to previous studies to account for the low S/N of our data and to avoid excessively large bins.

\begin{figure*}[h]
\centering
    \includegraphics[width=\textwidth]{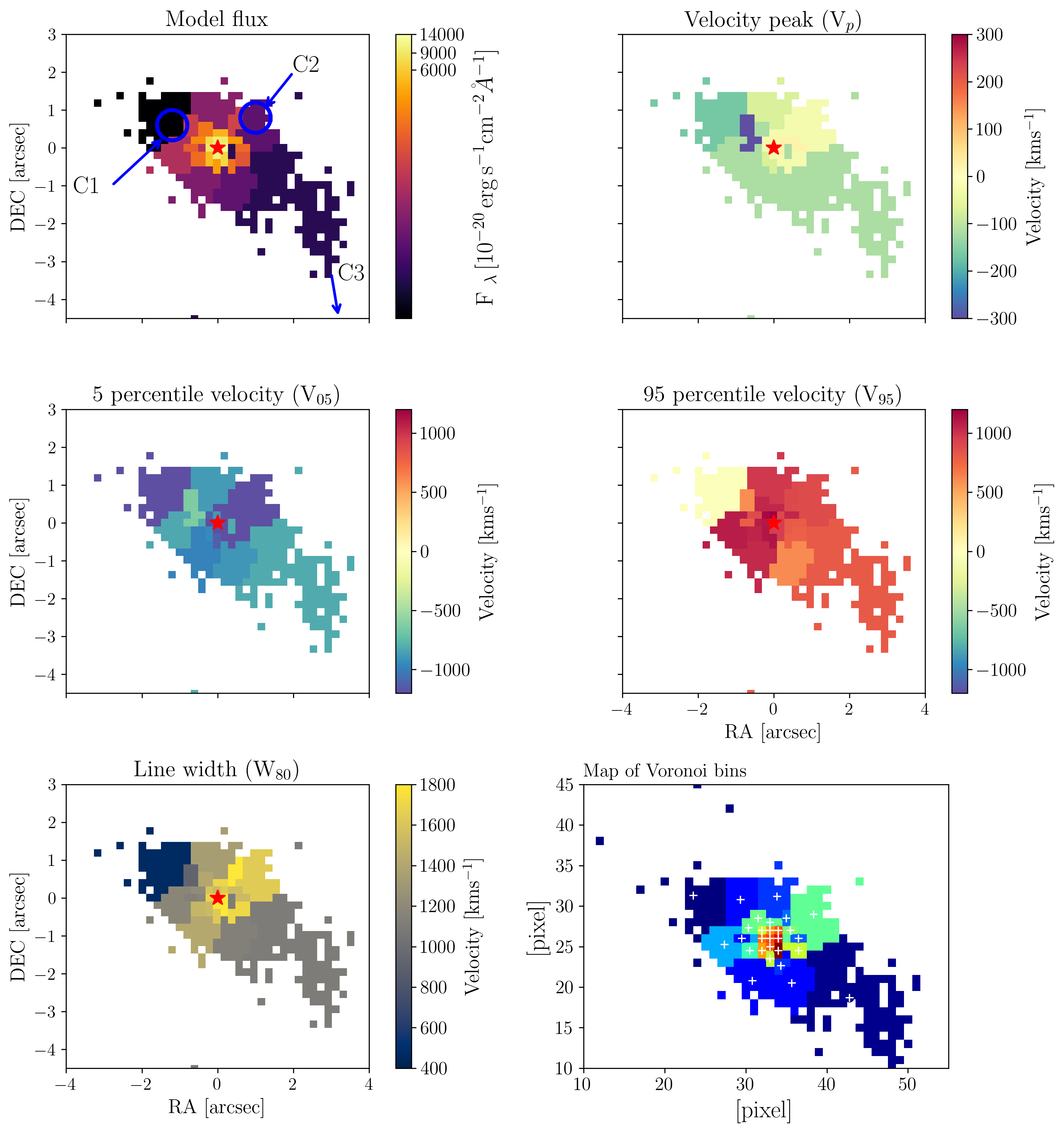} 
\caption[Three-Gaussian fitting of Voronoi bins.]{Results from three-Gaussian fitting of Voronoi bins. Top left panel: The flux map. Top right panel: the velocity peak. In the middle panel, we present the V$_{\rm{05}}$ and V$_{\rm{95}}$ maps indicating the velocities at the 5$^{\rm{th}}$ and 95$^{\rm{th}}$ percentile, respectively. Bottom left panel: The line width (W$_{\rm{80}}$). Bottom right panel: the Voronoi bins. }
\label{velocitymapsvor}
\end{figure*}

After the binning process, we applied three-Gaussian fitting procedures to the [O III]$\lambda$4959 and [O III]$\lambda$5007 doublet, similar to the one described in Appendix \ref{multiGaussian} for each Voronoi bin. The Voronoi bin map and the resulting flux and velocity maps are shown in Fig.$\;$\ref{velocitymapsvor}.

We compared our results from the three Gaussian fitting with Voronoi binning and without. The results for the velocity profiles are more or less similar giving the same velocity ranges. However, the velocity values in Voronoi binning maps are higher in the SW extended structure with V$_{\rm{05}}$ $\sim$\;-900\,$\; \rm{km\;s^{-1}}$ and V$_{\rm{05}}$ $\sim$\;700\,$\; \rm{km\;s^{-1}}$. This also has corresponding high value of velocity dispersion (W$_{\rm{80}}$) $>$\;800\,$\; \rm{km\;s^{-1}}$. The Voronoi bin size in this region is big because of the low signal-to-noise (S/N) and as a consequence the high kinematic values. The kinematics of our data are well-defined without the binning. Therefore the results in this study are based on the multi-Guassian fitting described in Sect. \ref{datafitting_spaxel}. 

\FloatBarrier

\end{appendix}

\end{document}